\documentclass[review]{elsarticle}

\usepackage{hyperref}
\journal{Journal of \LaTeX\ Templates}

\usepackage{amssymb}
\usepackage{mathtools}
\usepackage{graphicx}
\usepackage{xr}
\usepackage{booktabs}
\usepackage{cleveref}
\newcommand{\ER}{Erd\"os-R\'enyi }
\usepackage{amsmath}
\usepackage{algorithm}
\usepackage[noend]{algpseudocode}

\newcommand{\Lagr}{\mathcal{L}}
\DeclarePairedDelimiter\abs{\lvert}{\rvert}
\makeatletter
\let\oldabs\abs
\def\abs{\@ifstar{\oldabs}{\oldabs*}}
\begin{document}

\begin{frontmatter}

\title{Characterising brain network topologies: a dynamic analysis approach using heat kernels}
%%\tnoteref{mytitlenote}}
%%\tnotetext[mytitlenote]{Fully documented templates are available in the elsarticle package on \href{http://www.ctan.org/tex-archive/macros/latex/contrib/elsarticle}{CTAN}.}

%%\author{Elsevier\fnref{myfootnote}}
%%\address{Radarweg 29, Amsterdam}
%%\fntext[myfootnote]{Since 1880.}

\address[a1]{Department of Biomedical Engineering, Division of Imaging Sciences \& Biomedical Engineering, King's College London, London, UK}
\address[a2]{Stroke Division \& Massachusetts General Hospital, Harvard Medical School, J. Philip Kistler Stroke Research Center, Boston, USA}
\address[a3]{Centre for the Developing Brain, Division of Imaging Sciences \& Biomedical Engineering, King's College London, London, UK}

\author[a1]{A.W. Chung}
\author[a2]{M.D. Schirmer}
\author[a3]{M.L. Krishnan}
\author[a3]{G. Ball}
\author[a3]{P. Aljabar}
\author[a3]{A.D. Edwards}
\author[a1]{G. Montana\corref{mycorrespondingauthor}}
\cortext[mycorrespondingauthor]{Corresponding author}
\ead{giovanni.montana@kcl.ac.uk}

% \address[mainaddress]{Biostatistics and Bioinformatics, Division of Imaging Sciences \& Biomedical Engineering, King's College London, St. Thomas', London, UK}
% \address[secondaryaddress]{Division of Imaging Sciences \& Biomedical Engineering, King's College London, St. Thomas', London, UK}
% \address[tertiaryaddress]{Centre for the Developing Brain, King's College London, St. Thomas', London, UK}

\begin{abstract}

Network theory provides a principled abstraction of the human brain: reducing a complex system into a simpler representation from which to investigate brain organisation. Recent advancement in the neuroimaging field are towards representing brain connectivity as a dynamic process in order to gain a deeper understanding of how the brain is organised for information transport. In this paper we propose a network modelling approach based on the heat kernel to capture the process of heat diffusion in complex networks. By applying the heat kernel to structural brain networks, we define new features which quantify change in energy flow. Identifying suitable features which can classify networks between cohorts is useful towards understanding the effect of disease on brain architecture. We demonstrate the discriminative power of heat kernel features in both synthetic and clinical preterm data. By generating an extensive range of synthetic networks with varying density and randomisation, we investigate how heat flows in the networks in relation to changes in network topology. We demonstrate that our proposed features provide a metric of network efficiency and may be indicative of organisational principles commonly associated with, for example, small-world architecture. In addition, we show the potential of these features to characterise and classify between network topologies. We further demonstrate our methodology in a clinical setting by applying it to a large cohort of preterm babies scanned at term equivalent age from which diffusion networks were computed. We show that our heat kernel features are able to successfully predict motor function measured at two years of age (sensitivity, specificity, F-score, accuracy = 75.0, 82.5, 78.6, 82.3\%, respectively).\\

\end{abstract}

\begin{keyword}
brain connectivity networks \sep connectome \sep structural network \sep heat kernel \sep diffusion kernel \sep synthetic networks \sep preterm \sep developing brain \sep classification \sep motor function \sep diffusion MRI
%%\MSC[2010] 00-01\sep  99-00
\end{keyword}

\end{frontmatter}

%\linenumbers

\section{Introduction}
The human brain is a complex system of units (neurons) which interact with one another to process internal and external stimuli. In such complex systems, many features emerge due to their interaction and their global connections which can be analysed using graph theory.
The application of graph theory for investigating brain function and connectivity has been readily adopted by the neuroimaging community~\cite{bullmore_complex_2009, fornito_connectomics_2015}. As a mathematical model capturing relationships between interacting objects, a graph (or network) provides a simple abstraction of neural connectivity; reducing a complex system into a collection of \textit{nodes} (representing brain regions) which are connected by \textit{edges} representative of their relation. In diffusion magnetic resonance imaging (MRI) based structural networks, edges between brain regions signify their connection via an anatomical pathway from white matter tracts inferred using tractography. Edges may be assigned a \textit{weight} indicating the strength of the connection, such as the use of \textit{fractional anisotropy} as a measure of the pathway's structural integrity~\cite{jones_white_2013, fornito_graph_2013}. In functional MRI based networks, edges represent a measure of association in blood-oxygen-level-dependent signals across time, which reflect neuronal activity. The strength of this association may be indicative of how functionally related two regions are and is thus assigned as an edge weight~\cite{fornito_graph_2013}. As a branch of mathematics, graph theory offers a wealth of tools to describe networks in a rich form, making it an attractive framework for investigating brain organisation. For example, topological principles such as \textit{small-world} and \textit{rich-club} organisation have been found in many natural complex systems, including the brain~\cite{van_den_heuvel_small-world_2008, towlson_rich_2013, ball_rich-club_2014}. Networks with small-world architecture which may be characterised by both large clustering and short path lengths have been associated with efficient information transport~\cite{watts_collective_1998}. The rich-club can be seen as a highly inter-connected set of nodes which form a backbone of the network structure~\cite{heuvel_high-cost_2012} and its network-theoretical importance has been shown with respect to nodal specialisation, functional integration and resilience to "attacks"~\cite{colizza_detecting_2006, collin_structural_2014, mcauley_rich-club_2007}. Several other graph-theoretical measures have been investigated to describe these topological properties of the underlying brain connectivity, however, there is no consensus on which set of measures can be used to completely characterise the brain (for a review of commonly used measures see Rubinov and Sporns~\cite{rubinov_complex_2010}). 

The strength of a graph representation for brain characterisation lies in its simplicity. Graph topologies can be used to describe a number of \textit{neural mechanisms} which shape neural responses to a disease and its propagation through brain architecture~\cite{fornito_connectomics_2015}. The highly interconnected brain enables disease propagation across the organ via its axonal pathways~\cite{saxena_selective_2011, hirokawa_molecular_2010, perlson_retrograde_2010}. Thus disorders can have a pervasive effect on function and structure that is not necessarily localised to the region of insult or pathological onset. For example, stroke patients exhibit functional over-activation across brain regions that are remote from the vicinity of the lesion~\cite{rehme_cerebral_2013}. Another example is widespread neurodegeneration alongside disease progression in degenerative disorders such as Huntington's and Parkinson's diseases which are believed to have focal onset~\cite{tabrizi_biological_2009, goedert_100_2013}. An example neural response is \textit{dedifferentiation}, the recruitment of diffused, non-specific brain regions for task performance that is often observed in the ageing population~\cite{sleimen-malkoun_aging_2014} and schizophrenia~\cite{honey_functional_2005}. Another neural mechanism is \textit{compensation}, where functional activity is increased following an insult or in the early stages of a neurodegenerative disease and is frequently reported in multiple sclerosis~\cite{chiaravalloti_cognitive_2015} and Alzheimer's disease~\cite{elman_neural_2014}. As the spread and impact of these neural responses can be shaped by the underlying brain connectivity, network theory may provide quantitative descriptors of these mechanisms~\cite{fornito_connectomics_2015, schoonheim_network_2015}. Graph measures or features have thus been found to be associated with a number of neuropathologies~\cite{odish_dynamics_2015, lo_diffusion_2010, wang_altered_2009, pandit_whole-brain_2014}.

A main objective in neuroimaging studies is to elucidate how a specific disease affects the underlying network topology; gaining such an understanding then allows discrimination between patients and healthy controls. Identifying biomarkers of a disease would thus be useful for advanced diagnostic or predictive applications. The power of network-derived features for describing the human brain is evident by their increasing use in classification of neuroimaging data. Network classification involves categorising a network as belonging to a control or a disease population, or even to a subcategory in the case of spectrum disorders. Network classification requires the extraction of graph-based \textit{features} which are typically used as predictors in statistical classifiers. Studies have explored the discriminative power of network edges, revealing their promise in classifying a range of pathologies~\cite{zalesky_network-based_2010, shen_discriminative_2010,arbabshirani_classification_2013,richiardi_classifying_2012, prasad_brain_2015, rosa_sparse_2015}. Comparisons of graph metrics which characterise local and global topology as well as network principles have also been employed for classification purposes in major depressive disorder ~\cite{sacchet_support_2015}, Alzheimer's disease~\cite{prasad_brain_2015} and pre-school versus adolescent children~\cite{meskaldji_improved_2015}.\

The mechanisms by which neural impulses, or \textit{information}, propagate through the human brain network is limited by the finite propagation speed of the electro-chemical signals. Some network measures, such as shortest characteristic path length, do not incorporate the idea of information transport directly, but describe the structural (and static) connectivity profile while using shortest path lengths. However, given the propagative neural mechanisms discussed earlier, we hypothesise that capturing \textit{energy flow} through a network over 'time' could provide useful features for classification purposes. In this work, we propose the \textit{heat kernel} for capturing energy flow in a network. A heat kernel summarises the effect of applying a source of heat to a network and observing its diffusion process over 'time'. It encodes the distribution of heat over a network and characterises the underlying topological structure of the graph. This diffusion process, from which the heat kernel is the fundamental solution to, was widely used in image analysis for smoothing purposes~\cite{babaud_uniqueness_1986, perona_scale-space_1990}. This idea was later extended by applying the heat kernel on a graph representation of the image~\cite{zhang_graph_2008}. In the context of brain network analysis, only a handful of studies using heat kernels have been reported. They include an application on structural networks to investigate disease progression in Alzheimer's in which the eigenmodes of the heat kernel showed spatial similarity to the measured atrophy patterns from the grey matter volume~\cite{raj_network_2012}. Heat kernels have also been utilised to investigate the relationship between structural and functional networks~\cite{abdelnour_network_2014}. In these cases, analyses are performed with respects to a single heat kernel calculated with its time parameter fixed to a single value. In contrast, we propose an alternative approach where we make use of a time-series of heat kernels computed over a range of the time parameter. From this time-series, we derive features representative of energy transport which appear to capture salient network properties that can be used to discriminate between different network topologies. Recent and somewhat related work includes Mi\v si\'c et al.~\cite{misic_cooperative_2015}, who modelled the spread of local perturbations across brain networks and analysed the time it takes for a disturbance to a node to spread across the entire network. Via this dynamic model, the authors demonstrated the importance of structural hub regions as a backbone to facilitate rapid spreading and also relevant cooperative and competitive interactions between resting state networks. They found the structural network to support interactive relationships between functional modules.

Furthermore to our proposed heat kernel features, we present a framework for generating a baseline of synthetic networks to simulate brain networks of varying network densities and randomisation levels. With these synthetic networks, we investigate the changes in our heat kernel features with graph topology and demonstrate an association with small-world architecture. Subsequently, using linear discriminant analysis we show the ability of our heat kernel measures to classify between specific topologies. 
In addition, we apply our methodology to the problem of early detection of adverse neurological outcome that is common in children born very preterm (born at 32 weeks gestation or younger)~\cite{delobel-ayoub_behavioral_2009, edwards_developmental_2011}. Surviving preterm infants are susceptible to significant deficits in cognitive, behavioural and sensory development as well as long-term motor dysfunction with a high risk of cerebral palsy~\cite{marlow_motor_2007, back_brain_2014}. Associations between cognitive outcome and diffusion tractography features computed at term from premature neonates have been reported~\cite{van_kooij_neonatal_2012, duerden_tract-based_2015, ball_thalamocortical_2015}, demonstrating the advantage of imaging predictors for early diagnosis. The development of brain architectural features such as those proposed in our work may contribute towards understanding the neural mechanisms characteristic of functional deficits linked with prematurity. Obtaining predictors which are sensitive to neurodevelopmental outcome are also invaluable for early intervention and treatment planning to mitigate the impact of preterm birth. Thus we test the efficacy of heat kernel features computed from structual networks to be predictors of motor dysfunction in a cohort of preterms. By dividing the cohort into two groups depending on their mobility score, we demonstrate that our heat kernel features can predict the motor outcome of preterm babies scanned at term.\\

The rest of the paper takes the following format: in Section 2, we first detail our heat kernel methodology and synthetic network framework. This is followed by experimental settings for the synthetic networks and the clinical application on a premature cohort. Section 3 contains results of the experiments which are then discussed in Section 4. 

\section{Material and methods}

In this section we first provide the background and notations for graphs and heat kernels. We then describe our methodology and define the proposed heat kernel features. We next detail the framework for generating the synthetic networks, followed by descriptions of all experiments.\\

\subsection{Background}
\subsubsection{Graph notation}
Let a graph be represented as $G = (V,E)$ where $V$ is the set of $|V|$ nodes on which the graph is defined and $E \subseteq V\times V$ the corresponding set of edges. The adjacency matrix, $A$, is of size $|V| \times |V|$, where $A(u,v) = 1$ if an edge exists between nodes $u$ and $v$, and 0 otherwise. A weighted matrix, $W$, is defined as $W(u,v) = w_{uv}$ if $A(u,v) = 1$ and 0 otherwise, where $w_{uv}$ represents the corresponding edge strength. A diagonal strength matrix, D, is defined as $D(u,u) = deg(u) = \sum_{v \in V}w_{uv}$. The Laplacian, $\Lagr$, of $G$ is defined as $\Lagr = D - W$, and the normalised Laplacian is given by $\hat\Lagr = D^{-1/2}\Lagr D^{-1/2}$.\\

\subsubsection{The heat kernel}

Information transport within the brain can be described through the propagation of electro-chemical energy. The diffusion of energy or heat through a system is a known problem in physics. This diffusion process is described by the standard diffusion equation, a partial differential equation which expresses change in the density  of the diffusing material within any part of a system based on its flow. Its equivalence in the field of heat conductance is the \textit{heat equation}:

\begin{equation} \label{eq:heat_eq}
\frac{\partial H(t)}{\partial t} = -\hat\Lagr H(t),
\end{equation}

where the heat kernel, $H(t)$, is the fundamental solution. $H(t)$ can be viewed as describing the flow of energy through a network's edges at time $t$. The rate of flow is governed by $\hat\Lagr$ of the graph and its relationship with $H(t)$ has been widely studied in spectral graph theory~\cite{chung_spectral_1997, yau_lectures_1994}.

The heat kernel is a $|V| \times |V|$ matrix and can be computed by expressing $\hat\Lagr$ via its eigenspectrum, $\hat\Lagr = \Phi\Lambda\Phi^T$, where $\Lambda = diag(\lambda_{1}, \lambda_{2},...,\lambda_{|V|})$ is a diagonal matrix of eigenvalues ordered by increasing magnitude $(\lambda_{1} < \lambda_{2} < ... < \lambda_{|V|})$ and $\Phi = (\phi_{1},\phi_{2},...,\phi_{|V|})$ is a matrix of the corresponding eigenvectors as columns. The entries for the heat kernel between nodes $u$ and $v$ can be calculated as:

\begin{equation} \label{eq:HK_spect}
H(t)_{u,v} =  \Phi \exp[-\Lambda t]\Phi^{T} = \sum_{i=1}^{|V|} \exp[-\lambda_{i} t]\phi_{i}(u) \phi_{i}(v).
\end{equation}

The entry $H(t)_{u,v}$ represents the amount of heat initially placed on node $u$ that has reached node $v$ after time $t$. Thus $H(t)$ encodes the distribution of path lengths in a network such that the heat transference given by $H(t)_{u,v}$ occurs via all possible pathways connecting nodes $u$ and $v$. Should $A(u,v) = 1$, $H(t)_{u,v}$ will decay exponentially with the weight of the corresponding edge~\cite{zhang_graph_2008}. Intuitively, the stronger the connection between two nodes, the sooner heat will propagate between them. After the initial heat is applied to the network, $H(t)$ can be approximated by $H(t) \simeq I - \hat\Lagr t$ and the heat kernel depends on the local connectivity profile or topology of the graph. If $t$ is``large'', then $H(t) \simeq I - \exp[-\lambda_{2}t]\phi_2\phi_2^T $, where $\lambda_2$ is the smallest non-zero eigenvalue and $\phi_2$ the associated eigenvector (the Fiedler vector~\cite{fiedler_laplacian_1989}). Hence, the large time behaviour is governed by the global topology of the graph.

As an alternative to the numerical solution in Equation \ref{eq:HK_spect}, we can compute the heat kernel analytically by exponentiating $\hat\Lagr$ with time using the Pad\'e approximant~\cite{al-mohy_new_2009}:

\begin{equation} \label{eq:HK_eq}
H(t) = \text{exp}[-t\hat\Lagr].
\end{equation}

\subsection{Dynamic heat kernel features}

Based on the heat kernel computed from Equation \ref{eq:HK_eq}, several features can be extracted to represent the dynamic properties of the network. Of particular interest is the time when the relative change in heat transfer in the network has dropped below a given percentage. In this case the transference of the heat between consecutive time steps becomes small, compared to the amount of heat which has been transferred up to this time point. This means that the estimated heat kernel value at any given time point becomes "stable" with regards to small variations in time. We refer to the time it takes for the network to reach this level as the network's \textit{intrinsic time constant}, $t_c$. 

The intrinsic time constant $t_{c}(u,v)$ for an edge between nodes $u$ and $v$ is the maximal time when the relative percentage change in $H_{u,v}$ computed at consecutive time points falls below a percentage threshold, $s$: 

\begin{equation}
t_{c}(u,v) = t_{\max}:\abs{\frac{H(t+\Delta t)_{u,v}-H(t)_{u,v}}{H(t)_{u,v}}}_{t_{1}}^{t_{2}} < s,
\end{equation}

where $\Delta t$ is a time step within the range of $t_1 \leq t \leq t_2$.

During the process of energy transfer through a system, the exchange of heat will reach a maximum or peak. The maximal level, \textit{peak difference value}, of heat transfer and the time to reach this level, \textit{peak difference time}, following the introduction of heat into the network are two additional key features of this energy diffusion process. The peak difference value, $h_{peak}(u,v)$, is the largest difference in energy transferred between two consecutive time points and is given by

\begin{equation}
h_{peak}(u,v) = \max\abs{H(t+\Delta t)_{u,v}-H(t)_{u,v}}_{t_{1}}^{t_{2}}.
\end{equation}

The time at which this occurs is the peak difference time, with:

\begin{equation}
t_{peak}(u,v) = t:h_{peak}(u,v).
\end{equation}

$h_{peak}$ and $t_{peak}$ are representative of the maximal flow in energy occurring in the system. Table~\ref{tab:hk_measures} provides a reference of these heat kernel features.

\begin{table}[h!]
	\begin{center}
		\caption{Definitions of dynamic heat kernel features}
		\label{tab:hk_measures}
		\begin{tabular}{lll}
			\toprule
			\hline			
			Abbreviation & Measure & Definition\\
			\midrule
			$t_{c}$ & Intrinsic time constant & Time at which change in $H_{u,v}$\\ 
			& & drops below threshold\\	
			$h_{peak}$ & Peak difference value & Maximal exertion of energy transfer\\
			$t_{peak}$ & Time of $h_{peak}$ & Time that maximal exertion occurs\\
			$t_{c}^{g}$ & Global intrinsic time constant & A global metric of $t_{c}$\\
			$h_{peak}^{g}$ & Global peak difference value & A global metric of $h_{peak}$\\
			$t_{peak}^{g}$ & Global time of $h_{peak}$ & A global metric of $t_{peak}$\\
			\bottomrule
		\end{tabular}
	\end{center}
\end{table}

\subsection{A framework for synthetic networks}
In order to investigate the behaviour of the proposed heat kernel features in a controlled setting, we created synthetic networks on a topological spectrum between ordered (lattice~\cite{watts_collective_1998}) and random \ER networks~\cite{erdos_random_1959}. This allows us to systematically vary the synthetic networks' topology, and assess whether such changes can be detected by our proposed features. Computing a spectrum of synthetic networks with increasing randomness was first proposed by Watts and Strogatz~\cite{watts_collective_1998}. They introduced the term \textit{small-world architecture} to represent the network efficiency observed from their spectrum of synthetic networks. Given that the human brain has small-world properties makes their model an attractive method for generating synthetic networks for neuroimaging studies. Watts and Strogatz arranged nodes on the circumference of a circle, using a spatial embedding (or geometric position in space) to define the local neighbourhood of each node. Nodes were then connected to their nearest neighbours, where the number of connected neighbours governed the overall network \textit{density}, $d$. This resulted in a lattice network with a given network density. Increasing \textit{randomisation}, $p$, of such a lattice network is achieved by randomly rewiring an increasing percentage of edges. When all edges are rewired randomly, the resulting graph corresponds to an \ER random graph.

The method proposed by Watts and Strogatz, however, uses a 2D spatial embedding of the nodes to define the neighbourhood. In order for the synthetic networks utilised in this work to resemble human brain networks more closely, we use a 3D spatial embedding for our undirected synthetic networks. These synthetic networks comprise of two hollow, three-dimensional half spheres, each representing a brain hemisphere. Brain regions or nodes are randomly defined on the surface using Poisson disk sampling~\cite{bridson_fast_2007}. This sampling technique defines regions based on a distance threshold between region centres and has been used to define regions in neuroimaging studies of the human brain~\cite{ball_rich-club_2014, schirmer_parcellation-independent_2014}. Advantages of this technique are that regions are tightly packed yet centres are no closer to each other than this minimum distance, thereby creating an (approximately) uniformly spaced grid of nodes of similar size on the half spheres. With the nodes equally distributed and spatially delineated in space, lattice-like connectivity is defined by connecting each node to its spatially adjacent neighbours, resulting in a spatial adjacency matrix $A_{sp}$.

\begin{figure}[h]
\centering
\includegraphics[width=1 \textwidth]{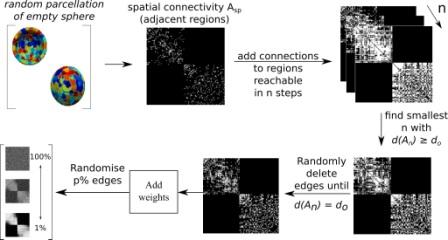}
\caption{A framework for generating synthetic networks with varying density, $d$, and randomisation levels, $p$. Two surfaces of half spheres, each ‘representing’ a brain hemisphere, are randomly parcellated into 100 regions each. Parcellation was achieved using Poisson disk sampling, which defines spatial adjacencies and represents a 3D lattice network $A_{sp}$. An observed density $d_o$ is achieved by initially interconnecting all regions, which are a distance $n$ apart, based on $A_{sp}$, until $d_o$ is exceeded or reached (we call this computed density $d(A_n)$). If $d(A_n) \geq d_o$, edges are randomly deleted until $d(A_n) = d_o$. Subsequently, weights are assigned randomly from a normal distribution($\mathcal{N}$(1,0.25)) and $p\%$ of the edges are randomised.}
\label{fig:surr_pipeline}
\end{figure}

In our surrogate experiments, we explore topologies over a range of both network density and randomisation $(d,p)$. The graph density is controlled by varying the ``depth'' to which a node is connected beyond its immediate spatial neighbour. This is achieved by calculating the shortest path distances $\Lambda$ in $A_{sp}$ and edges are added to each node's  $n^{th}$-neighbourhood, where $n$ is the length of path-ways defined by $\Lambda$. In order to match the density of the surrogate networks to a particular density percentage, $d_{o}$, the lattice-like connectivity of the synthetic network is increased until its density $d$ either matches of exceeds $d_{o}$. If $d_{o}$ is exceeded, edges in the network are randomly deleted, until $d=d_{o}$. The process of achieving a specific density level can be repeated multiple times via this random deletion of edges, each time generating a density matched network with a different connectivity profile. In our work, we generate one spatial adjacency matrix using Poisson disk sampling, from which a set of density matched networks can be created. Edge weights are then randomly drawn from a normal distribution $\mathcal{N}$(1,0.25), resulting in a density matched, weighted lattice network with similar degree distributions. Networks with increasing randomisation are then created by randomly rewiring $p\%$ of the edges). Figure~\ref{fig:surr_pipeline} illustrates our framework for generating a spectrum of synthetic networks.\\

\subsection{Computing global features for analysis}

\begin{figure}[h]
\centering
\includegraphics[width=0.4 \textwidth]{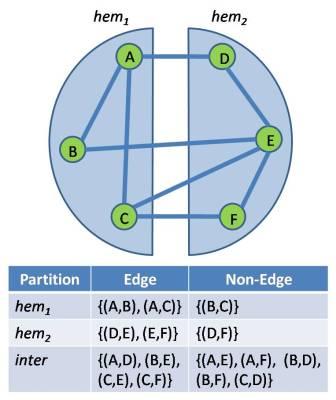}
\caption{Example illustration of how partitions are defined. Nodes A to F exist in hemispheres, \textit{$hem_{1}$} and \textit{$hem_{2}$}, and edges are defined by blue lines. Set memberships for each of the six partitions for all potential pairwise connections in this network are listed in the table.}
\label{fig:partitions}
\end{figure}

For analysis, each heat kernel feature is converted into a global measure by first grouping edges into \textit{partitions}. First, an entry $A(u,v)$ is categorised by whether nodes $u$ and $v$ lie within the same hemisphere, \textit{$hem_{1}$, $hem_{2}$} (for hemisphere 1 and 2 respectively), or in different hemispheres, \textit{$inter$}. Within these three categories, each $A(u,v)$ is further distinguished as belonging to an edge or non-edge partition (\textit{$edge$, $\neg edge$}) depending on whether $A(u,v) = 1$ or $0$, respectively. Figure~\ref{fig:partitions} is an illustration of how partitions are defined. Median heat kernel features are calculated for each of these six partitions. For \textit{$edge$, $\neg edge$} partitions, a global mean is determined from their respective $hem_{1}$, $hem_{2}$ and $inter$ partitions. The contribution of the intra-hemispheric partitions are first summarised. This is because the initial connectivity profiles for each hemisphere are created independent of each other and randomisation occurs without priors, resulting in similar profiles. As an example, $\neg edge$ global $h_{peak}$ is calculated as follows using the median $h_{peak}$ from all $\neg edge$-related partitions: $h_{peak}^g = ((hem_{1}+hem_{2})/2+inter)/2$.

\subsection{Synthetic networks experimental settings}

\noindent
\textit{Experimental Parameters}\\
\indent
Synthetic networks were constructed with 200 regions (100 regions for each hemisphere) across a range of $d = [10,11,...,50\%]$, each with randomisation percentages of $p = [1,2,...,100\%]$. Fifty replicates were created for each combination of $(d,p)$. Heat kernels were calculated with $t = [0.05, 0.1, ..., 15.0]$
 (i.e. $\Delta t = 0.05$, resulting in 300 heat kernels for each network), and $t_{c}$ was computed with a threshold of $s = 2\%$. Median heat kernel features were calculated for each partition to obtain global measures.\\ 

\begin{table}[h!]
	\begin{center}
		\caption{List of global heat kernel feature sets used for classification}
		\label{tab:features_list}
		\begin{tabular}{l}
			\toprule
			\hline			
			Feature Sets\\
			\midrule
			1) $t_{c}^{g}$ \\	
			2) $h_{peak}^{g}$ \\
			3) $t_{peak}^{g}$ \\
			4) $t_{c}^{g}$ and $h_{peak}^{g}$ \\
			5) $t_{c}^{g}$ and $t_{peak}^{g}$ \\
			6) $h_{peak}^{g}$ and $t_{peak}^{g}$ \\
			7) $t_{c}^{g}$ and $h_{peak}^{g}$ and $t_{peak}^{g}$ \\
			\bottomrule
		\end{tabular}
	\end{center}
\end{table}

\noindent
\textit{Synthetic network classification}\\
\indent
We hypothesise that networks with small-world topology will be the most effective at distributing heat, as they are believed to be most efficient for information processing and learning ~\cite{simard_fastest_2005, bassett_adaptive_2006}. To investigate this, we determined the classification performance of heat kernel features to distinguish between the topology that exhibited 'greatest efficiency' and each of the remaining $(d,p)$ topologies that were generated. Specifically, a representative most efficient network, $sw(d,p)$, was determined for a subset of the densities, $d = [15, 20, 30, 40\%]$. For each $d$, the median $t_{c}^g$ (where the median was calculated across the fifty replicates at each $(d,p)$) were smoothed across the range of $p$ using a Savitzky-Golay smoothing filter~\cite{savitzky_smoothing_1964}. The minima and its corresponding randomisation percentage of this curve were identified for each $d$ to be its representative $sw(d,p)$ network.
A total of seven feature sets comprising of combinations of the three global heat kernel measures as detailed in Table~\ref{tab:features_list} were investigated. Linear discriminant analysis (LDA) ~\cite{pedregosa_scikit-learn:_2011} was performed to classify between $sw(d,p)$ and all remaining $(d,p)$ topologies for each feature set and for $edge$ and $\neg edge$ partitions separately. Prior to classification, all features were standardised by centering to the mean with unit variance. A stratified, 10-fold cross-validation scheme was used to assess classifier performance, and classifier performance measures of sensitivity, specificity, F-score and accuracy were recorded. See ~\ref{app:alg_synthetic} for a pipeline of the classification experiments.\\

\subsection{Application - Preterm cohort}

To demonstrate our methodology on neuroimaging data, we computed heat kernel features on a preterm cohort to investigate whether our features can be used to classify between infants with poor and normal motor ability.

\subsubsection{Data and image preprocessing}

Ethical permission for this study was granted by the Hammersmith, Queen Charlotte's and Chelsea Research Ethics Committee and written parental consent was obtained for each infant.\\

\noindent
\textit{Demographics and motor score}\\
\indent
290 infants were scanned at term equivalent age and all showed no evidence of focal abnormality on conventional MRI. Each subject had a neurodevelopmental assessment around 2 years corrected age ($20.18 \pm 8.2$ in mean(months.days) $\pm$ stdev (days)) using the Bayley-III test~\cite{bayley_bayley_2006}. A composite motor score was calculated and normalised with a mean of 100 and a standard deviation of 15. Two mobility groups were defined such that subjects with a composite motor score of 85 or less (i.e., 1 standard deviation below the mean for neuromotor function) were considered to have adverse mobility (n = 55, born at $28.28 \pm 2.25$ weeks gestational age (GA), scanned at $43.65 \pm 3.70$ weeks (scan age, SA)) and those greater than 85 to be controls with normal motor function (n = 233, born at $30.01 \pm 2.23$ weeks GA, scanned at $42.45 \pm 1.84$ weeks).\\

\noindent
\textit{MRI acquisition}

T1-, T2-, and diffusion-weighted MRI data were acquired on a Philips 3 T system (Philips Medical Systems, Netherlands) using an eight-channel phased array head coil. T1-weighted parameters were: repetition time (TR) = 17 msec; echo time (TE) = 4.6 msec; flip angle = $13^{\circ}$; field-of-view (FOV) = 210 $\times$ 210 mm$^2$; matrix = 256 $\times$ 256; voxel size = 0.82 $\times$ 0.82 $\times$ 0.8 mm$^3$. T2-weighted fast-spin echo parameters were: TR = 14.73 sec; TE = 160 msec; flip angle = $90^{\circ}$; FOV = 220 $\times$ 220 mm$^2$; matrix = 256 $\times$ 256; voxel size = 0.86 $\times$ 0.86 $\times$ 2 mm$^3$. Single shot diffusion-weighted echo-planar imaging was applied in 32 non-collinear directions with parameters: TR = 7536 msec; TE = 49 msec; flip angle = $90^{\circ}$; FOV = 224 $\times$ 224 mm$^2$; matrix = 128 $\times$ 128; voxel size = 1.75 $\times$ 1.75 $\times$ 2 mm$^3$; \textit{b}-value = 750 s$^{-1}$ mm$^{2}$.\\

\noindent
\textit{Image processing}\\
\indent
The structural T2 images were segmented using Automated Anatomical Labeling~\cite{tzourio-mazoyer_automated_2002} to parcellate the cortex in each scan. All sets of cortical ROI were transformed from T2-space into diffusion space using non-rigid T2-to-B0 registration using the IRTK software package (https://biomedia.doc.ic.ac.uk/software/irtk/). Prior to processing, all datasets were visually assessed for motion artefacts. Diffusion data were pre-processed using the FMRIB Software Library (FSL) Diffusion Toolkit (FDT; www.fmrib.ox.ac.uk/fsl/). For each set of cortical target regions, 1000 streamlines were propagated per seed voxel using a modified version of ProbtrackX~\cite{behrens_probabilistic_2007} in which \textit{integrated} anisotropy~\cite{robinson_identifying_2010} was used to define the weights of the structural connectivity between brain regions.

\subsubsection{Application experimental settings}

We extracted features from heat kernels calculated from each infant's weighted connectivity matrix for $t = [0.05, 0.1, ..., 15.0]$. $t_{c}$ was calculated for a range thresholds $s = [1, 2, ...5]$. In this application, partitions were simplified to $edge$ or $\neg edge$ without information on hemispheric or inter-hemispheric membership. Analyses was not partitioned by hemisphere due to the nature of brain asymmetry where equal contribution from measures from either hemispheres cannot be assumed as in the synthetic model. For example, $t^{g}_{peak}$ for an $edge$ partition was the median $t_{peak}$ from all edges where $A(u,v) = 1$ and the equivalent global measure for the $\neg edge$ partition was the median from pairwise connections where $A(u,v) = 0$.

The same feature sets as in Table~\ref{tab:features_list} were used for Gaussian Na\"{i}ve Bayes (GNB) classification. Including the five thresholds for $t_c$ (i.e. $t_{c1\%}$ to $t_{c5\%}$), we tested 23 feature sets in total. A stratified 10-fold cross-validation strategy was employed, and repeated five times. All features were standardised by centering to the mean with unit variance. Classifier performance measures of sensitivity, specificity, F-score and accuracy were calculated for each repetition and their averaged values are reported. The above classification was performed twice more, with GA and then SA each linearly regressed from the features. See ~\ref{app:alg_ePrime} for a pipeline of the classification experiments. As a comparison, standard network measures were also calculated for each subject and similarly used to classify between motor ability groups. These measures were: average characteristic path lengths, average edge betweenness centrality, average clustering coefficient and global efficiency (computed using BCT; brain-connectivity-toolbox.net). Characteristic path length was also defined for $edge$ and $\neg edge$ partitions. Feature sets combining these standard network measures for classifying are listed in ~\ref{app:standard_featuresets}.

\section{Results}

\subsection{Synthetic networks}

\begin{figure}[h!]
\centering
\includegraphics[width=1 \textwidth]{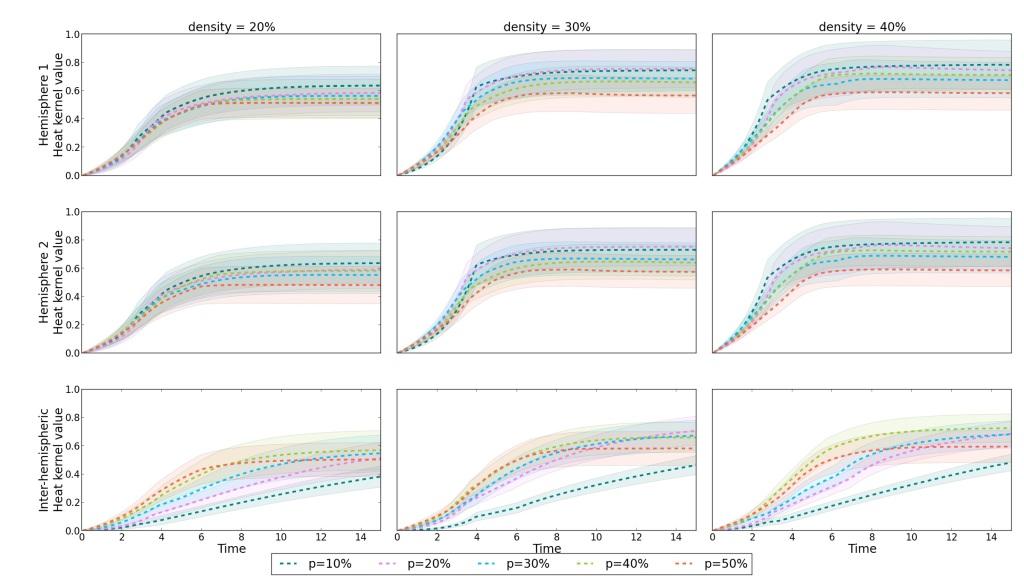}
\caption{Change in average heat kernel values for each $\neg edge$ partition with time of a single synthetic network with densities $d$ = [20, 30, 40\%] at a range of randomisation percentages, $p$. Mean and standard deviation are over all heat kernel `edges' within each partition.}
\label{fig:surr_HKwTime}
\end{figure}

Figure~\ref{fig:surr_HKwTime} illustrates the amount of heat captured in synthetic networks with the time parameter. Specifically, it plots the mean heat kernel values, $H(t)_{u,v}$, in $\neg edge$ partitions versus $t$ for a selection of topologies. It can be observed that the slope and shape of the curves vary depending on the partition: inter-hemispheric connections exhibit a more gradual incline in heat transfer with time, taking longer to stabilise than pathways between nodes which are within the same hemispheres ($hem_1$ and $hem_2$ regions). Also to note is the similarity in the trends plotted for each hemisphere. In addition the larger the density of the network, the larger the values of $H(t)_{u,v}$. That is, a more interconnected network with more edges enables a greater proportion of heat to reach node $v$ from $u$, as more heat can be 'stored' within the network.

The effect of randomising the synthetic networks affects heat transfer differently to changes in $d$. For \textit{$hem_{1}$} and \textit{$hem_{2}$}, increasing $p$ generally leads to a reduction in heat transfer (within time) whereas inter-hemispheric heat transfer increases. Edge partitions overall revealed a sharp or steady increase to similarly large heat kernel values irrespective of $d$ and $p$ (see~\ref{app:HK_edge}). This is because two nodes directly connected by an edge will have a consistent heat transfer between them compared to node pairs in $\neg edge$ partitions. For simplicity, all further synthetic network results will be reported for $\neg edge$ partitions only (however $edge$-based results can be found in the Appendix).\\

\begin{figure}[h!]
\centering
\includegraphics[width=1 \textwidth]{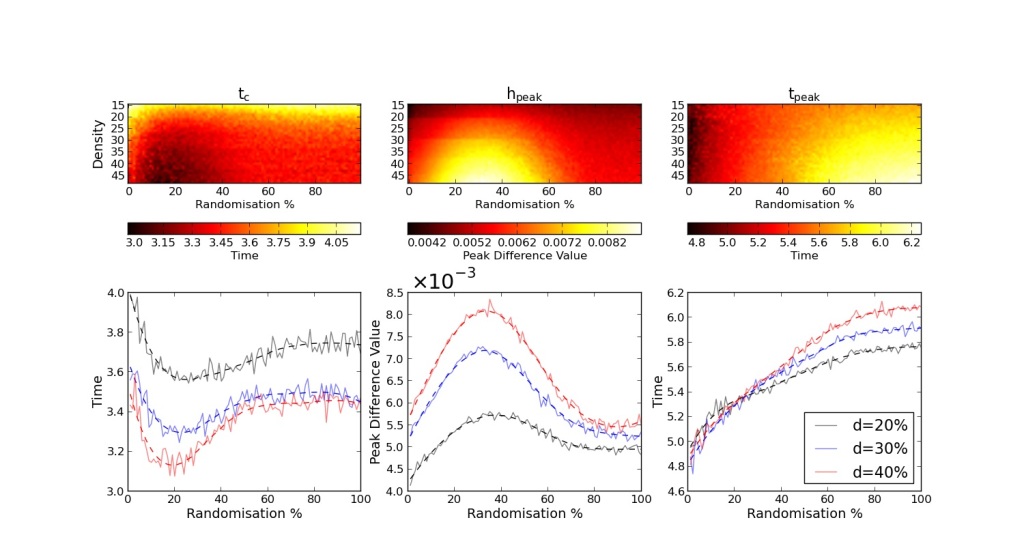}
\caption{Top row) Topology maps for each global heat kernel feature in the $\neg edge$ partition. Each pixel in the map represents the mean (taken across all fifty replicates) heat kernel feature for a synthetic network topology with density, $d$ ($y$-axis), and randomisation percentage $p$ ($x$-axis).  Bottom row) Example densities (i.e. rows) from the above corresponding topology maps depicting change in heat kernel feature with increasing randomisation percentage. Dashed line plots are interpolated global features across $p$ using a Savitzky-Golay filter.}
\label{fig:surr_gridmaps}
\end{figure}

The top row in Figure~\ref{fig:surr_gridmaps} contains topology maps showing global heat kernel features across synthetic replicates for all $(d,p)$. That is, each pixel in the map represents a $(d,p)$ topology, and contains the average global heat kernel feature over fifty $(d,p)$ networks. Each graph in the bottom row of Figure~\ref{fig:surr_gridmaps} plots the values in the corresponding topology map above for a selection of densities. For the intrinsic time constant increasing density leads to an overall decrease in $t_{c}^{g}$. The effect of randomisation percentage reveals there is a region of $p$ in which $t_{c}^{g}$ is minimal. Overall, this can be visualised as a semi-circle centred around $p \approx 20\%$ in the corresponding topology map. Given the variations in the realisations of these networks, the minima may not be well defined, however, there appears to be a dependency on $d$ in that the greater the network density, the lower the corresponding $p$ where $t_{c}^{g}$ is minimal ($t_{c}^{g}$ line plot in Figure~\ref{fig:surr_gridmaps}). This suggests that for a given density, the level of network randomisation when $t_{c}^{g}$ is smallest may represent a network most efficient for energy transport as heat exchange begins to 'stabilise' earlier (compared to networks at other $p$). The range of $p$ identified by $t_{c}^{g}$ also revolves around low levels of randomisation percentages which have been associated with small-world topology~\cite{watts_collective_1998}.\\
$h_{peak}^{g}$ varies with $d$ and $p$ in an opposite manner to $t_{c}^{g}$. In contrast, $t_{peak}^{g}$ increases with network randomisation, with greatest differences between densities at large $p$. Interestingly, irrespective of density, $t_{peak}^{g}$  converges around the same network randomisation level associated with small-world topology (i.e. approximately where $p \approx 20\%$ in $t_{peak}^{g}$ line plot in Figure ~\ref{fig:surr_gridmaps}). Results for the $edge$ partition can be found in ~\ref{app:HK_globalChange_edge}.

\begin{figure}[h!]
\centering
\includegraphics[width=1 \textwidth]{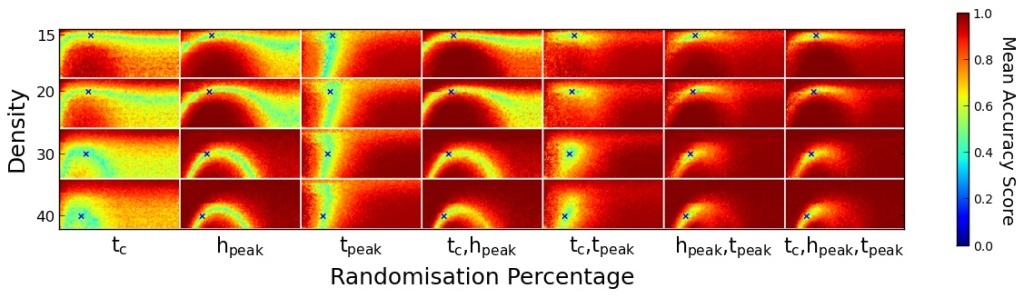}
\caption{Mean accuracy maps, $\neg edge$ partition - LDA classification accuracy scores for $sw(d,p)$ for $d=[15,20,30,40\%]$ versus all other topologies, for the seven feature sets tested. $sw(d,p)$ for each $d$ are indicated by an 'x'.}
\label{fig:surr_acc_gridmaps}
\end{figure}

Figure~\ref{fig:surr_acc_gridmaps} reports the accuracy score of LDA at classifying $sw(d,p)$ versus all other topologies, i.e. each pixel at position $(d,p)$ is coloured by the accuracy score ($\in [0,1]$)  from classifying its synthetic network against $sw(d,p)$. Plots for sensitivity, specificity and F-score show similar behaviour to accuracy and can be found in~\ref{app:sw_class}. These maps reveal important information about each of our heat kernel features. Three general regions can be identified from the $t_c$ and $h_{peak}$ maps: the 'ribbon' of topologies with less than 20\% density, the topologies with high randomisation and the 'small-world' semi-circle (see~\ref{Afig:example_topological_regions} for an illustration). The intrinsic time constant varies sharply between the three regions however appears homogeneous within each region. This suggests $t_{c}^{g}$ can distinctly capture these three general network features in our synthetic networks. $h_{peak}^{g}$ accuracy varies (with the increasing densities tested) within the small-world region as distinct layers of arches. The accuracy of $t_{peak}^{g}$ appears homogeneous across densities for a narrow randomisation range around $sw(d,p)$. To confirm that these results are not dependent on choosing to classify against a candidate small-world topology, we repeated this experiment on a candidate \textit{random} topology, $random(d,p)$ (see~\ref{app:rand_class}). We found the results based on $random(d,p)$ identified the same three regions presented here. Furthermore, $t_{peak}^{g}$ revealed a similar stratified representation in the high randomisation region. For both $sw(d,p)$ and $random(d,p)$ results, the combination of $h_{peak}$ and $t_{peak}$ perform similarly to that from combining all three features. Compared to feature set 6 (Table~\ref{tab:features_list}) combining all three features resulted in only a 5.73\% gain in number of topologies classified with accuracy $>80\%$. For completeness, $sw(d,p)$ classification results on the $edge$ partition can be found in~\ref{app:edge_sw_class}.

\subsection{Application - Preterm cohort}

\begin{figure}[h!]
\centering
\includegraphics[width=0.5 \textwidth]{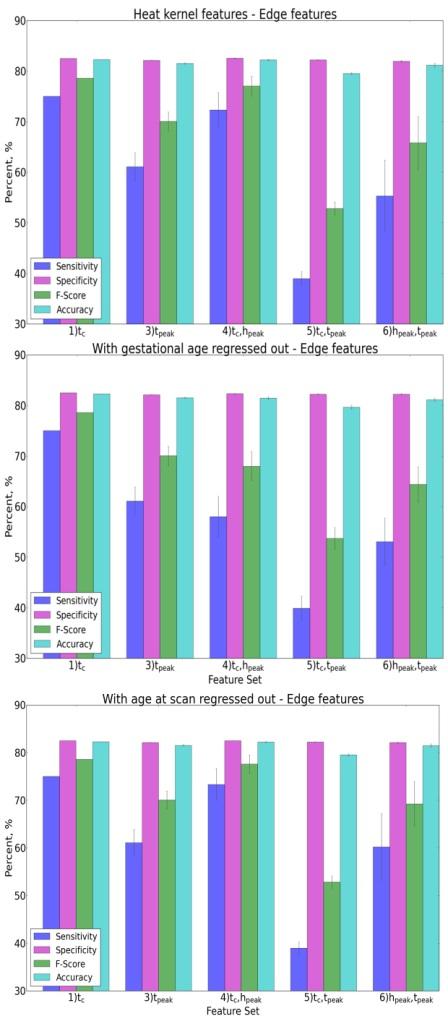}
\caption{GNB classification performance by feature set using heat kernel features calculated from the $edge$ partition (top row). Each bar represents the average score over five iterations of stratified 10-fold cross-validation. Error bars indicate standard deviation across iterations. Middle row are classification results using feature sets with GA regressed out. Bottom row contain results from classification repeated on feature sets with AS regressed out. Feature sets involving $t_c$ are computed with threshold $s=2$}
\label{fig:ePrime_subset_HK_edge}
\end{figure}

Results classifying the preterm cohort by motor function are presented in Figure~\ref{fig:ePrime_subset_HK_edge}.  Averaged classification performances across the five cross-validation iterations are plotted for a selection of feature sets tested from the $edge$ partition. For feature sets involving $t_c$ only results for $s=2\%$ are shown in the figure for simplicity. Bar plots for all feature sets as well as those for the $\neg edge$ partition can be found in~\ref{app:ePrime_class_non-edge}. All classification performance measures are listed in ~\ref{app:ePrime_class_tables}.  Heat kernel measures are accurate for classifying between preterm with normal and adverse motor function across a variety of feature sets, particularly feature sets 1) and 4) (Figure~\ref{fig:ePrime_subset_HK_edge}, top row). Regressing GA had the greatest effect on sensitivity in feature sets that combined $h_{peak}$ information, specifically feature sets 4) and 6), which decreased by 19.8\% and 4.1\%, respectively (Figure~\ref{fig:ePrime_subset_HK_edge}, middle row). Accounting for SA made little difference to classification performance (Figure~\ref{fig:ePrime_subset_HK_edge}, bottom row).  In general, $t_{c}$ performance scores remain the most stable of all feature sets after accounting for GA and SA (see ~\ref{Afig:ePrime_fullFeatureSets_edge}). From Figure~\ref{fig:ePrime_subset_HK_edge}, ${t_c}$ performs the best out of all features sets tested with sensitivity, specificity, F-score, and accuracy of 75.0, 82.5, 78.6, 82.3\%. For the $\neg edge$ partition, $h_{peak}$ classification performance is the best in term of sensitivity, with sensitivity, specificity, F-score, and accuracy of 65.6, 83.0, 73.2, 82.2\% (Appendix~\ref{Atab:nonedge_HK_GA}). For comparison, classifying using standard network measures did not perform as well, with average performance scores of 39.1, 83.3, 52.4, 73.3\% (for sensitivity, specificity, F-score, and accuracy. See Appendix~\ref{Atab:standard_measures} for full results).
\section{Discussion}

In this paper we proposed new heat kernel features which capture energy flow in a system to characterise and discriminate between network topologies in synthetic and \textit{in-vivo} data. We demonstrated the efficacy of the heat kernel for classifying structural networks using features which incorporate change in heat flow in networks over time as analysed by partitions. In addition, we also presented a new framework for building 3D embedded synthetic networks with a range of topologies for investigating our features.\\

\noindent
\textit{Understanding heat flow via analysis by partitions}

The heat kernel entry, $H(t)_{u,v}$, relates to the energy that has arrived at node $v$ from node $u$ at time $t$. This heat transference accounts for all possible paths which connect $u$ and $v$, not just that of the shortest paths within the network. Given the neural mechanisms for disease propagation and the idea that many neurological and psychiatric disorders can be described as dysconnection syndromes~\cite{fornito_connectomics_2015, catani_rises_2005}, understanding network features of energy flow partitioned according to directly (or indirectly) connected pairwise regions may provide insight into brain organisation. Thus by dividing our analysis into $edge$ and $\neg edge$ partitions, we can understand network topology in two ways: Firstly, features in the $edge$ partition not only inform us of the connection strength between any two directly connected nodes, but all other possible pathways between them throughout the network are also captured. Thus heat kernels have the advantage of not placing assumptions which constrain energy to flow only along the edge which connect node pairs. When applied to pathology, investigating $edge$ partitions 'locally' (i.e. localised to regions affected by the disease) could reveal the effect of an injury on the heat exchanged on a damaged connection and the consequence of this via heat flow through alternative, potentially compensatory, routes. Secondly, measures from the $\neg edge$ partition are informative of the underlying global connectivity of the brain. The heat measured in $\neg edge$ partitions capture 'communication' between indirectly connected node pairs. Thus $\neg edge$ partitions are not only indicative of the shortest path lengths between node pairs but are inclusive of all possible routes. An additional motivation to investigate $\neg edge$ partitions is because two regions which are not directly connected may still be mutually involved with a neural processing task~\cite{buckner_organization_2011, vincent_intrinsic_2007}.

The above intuitions can aid our understanding of how the heat kernel captures energy transport in relation to network topologies which vary with density and randomisation as modelled by our synthetic networks. In the context of $\neg edge$ partitions, larger density networks led to greater heat transference. This may be explained by the increase in number of connections and capacity of the network to store heat. With respect to network randomisation, the effect of increasing $p$ on energy transport depended on the partition. For within-hemispheric pathways, increasing $p$ reduced heat transfer as the chances of within-hemispheric edges being randomly assigned to between-hemisphere connections increased. Subsequently, for between-hemispheric pathways the opposite occurred - increasing $p$ led to larger heat transfer. This may be due to the increasing chance of inter-hemispheric connections being assigned thus making it easier for energy to traverse hemispheres more efficiently. In comparison, heat transference in $edge$ partitions exhibited similar trends with randomisation, but possessed slopes with varying degrees of change (~\ref{app:HK_edge}). The difference in heat transfer in relation to the corresponding $\neg edge$ counterpart still performed similarly well when classifying networks (~\ref{app:edge_sw_class}). These trends suggest that changes in heat kernel values with time can provide an indication of the underlying network structure in networks, particularly when analysed by partitions.\\

\noindent
\textit{Interpretation of heat kernel features}

We proposed three heat kernel features that can quantify important properties of information transport in a network. These features represent energy transference beginning to stabilise in the system ($t_{c}$), and a notion of when a peak in information flow occurs ($h_{peak}$ and $t_{peak}$). 

Small-world organisation is believed to be associated with efficient information propagation and as a topology, exists between that of an ordered, lattice network and an \ER network. By gradually increasing the randomisation percentage in a lattice network, Watts and Strogatz showed the emergence of small-world topology by rewiring only a small proportion of the edges, i.e. at low randomisation percentages~\cite{watts_collective_1998}. We found a similar trend within our surrogates as measured by our heat kernel features, particularly in the $\neg edge$ partition. Specifically, the time at which the relative heat transfer between consecutive time points "stabilised" occurred earliest in the lower ranges of $p$. Our results also indicate a range of network topologies in this region, which may be particularly efficient for information propagation. In addition, as connection density in a network increased (that is, the network gets closer to being fully connected), global efficiency also increased as additional edges lead to easier information transport between nodes~\cite{bullmore_brain_2011}. The heat kernel features capture this characteristic with decreasing/increasing values of $t_{c}$ and $h_{peak}$, respectively, with increasing density.

A parameter to consider is that of the threshold, $s$, when computing $t_{c}$. In our preterm application, classification performance was high across the range of $s$ in the $edge$ partition. For the $\neg edge$ partition, classification performed better at larger thresholds. Rather than suggesting insensitivity in $s$ for calculating the intrinsic time constant, there may be value in varying the degree in which to measure $t_c$. There are a number of factors on which $s$ may be dependent upon. $s$ is related to the resolution of the time steps used to compute the heat kernels, $\Delta t$. The choice of $\Delta t$ in turn may be dependent on the size of the network. This comes from the intuition that larger networks may need more time for energy to propagate through all nodes and an adjustment in $\Delta t$ may be necessary to capture the heat transfer with appropriate detail.

As an example application, we used heat kernel features for classifying between preterm infants with normal and adverse motor function. MRI-based features such as white matter injury (WMI), intraventricular haemorrhaging (IVH) or diffusion MRI measures of white matter tract integrity from infants scanned near term-equivalent age have been shown to be associated with developmental outcome~\cite{chau_abnormal_2013, ball_thalamocortical_2015, brown_prediction_2015}. Brown \textit{et al} classified preterm infants by motor score and showed that a combination of standard global network measures from diffusion tractography, WMI, IVH and GA achieved sensitivity, specificity and accuracy scores of 66, 79, 72.3\%, respectively~\cite{brown_prediction_2015}. We were able improve upon this by classifying with performance scores of 75.0, 82.5, 82.3\%, respectively. In addition, heat kernel features faired better than standard network measures in classifying our cohort. These results demonstrate the potential of our novel features based on energy propagation as extracted from heat kernels to predict preterm motor outcome at two years using structural networks.\\

\noindent
\textit{Future work}

The heat kernel methodology presented can be extended to resting-state functional networks. Traditional network measures which rely on paths and path lengths may not be suitable for functional networks as an interpretation of information transport, which is more appropriate for networks based on physical, anatomical connections. With several examples of strong, functional connections between regions with no direct anatomical link~\cite{buckner_organization_2011, vincent_intrinsic_2007}, a diffusion model such as the heat kernel which incorporates information as a composite of many pathways across the entire network could be more applicable. With increasing evidence to support the integrated and dynamic coordination of resting-state networks (or modules) for cognitive processing~\cite{fornito_competitive_2012, hellyer_control_2014, kragel_functional_2015}, our heat kernel features may shed light on module interdependence by investigating ``$\neg edge$'' partitions (inter-modular connections). 
Another interesting extension is to use our features to investigate the interplay between anatomical and functional networks. The associations between each of these biological networks with respects to neurological diseases and cognitive processing is often studied, however a deeper understanding of the structural underpinnings which lead to functional activation is necessary for a complete picture of brain architecture. Abdelnour \textit{et al.} ~\cite{abdelnour_network_2014} have shown that the heat kernel computed from a structural network bears similarity to the corresponding empirically computed functional network. Given the ability of the heat kernel features presented in this work to capture global network properties through energy propagation, it may be possible to demonstrate a link between features computed from structural and functional networks.

\section{Conclusion}

In this paper we presented new heat kernel features which capture energy propagation through structural networks. With a series of synthetic networks we explored heat diffusion in varying topologies by partitioning connections in the graph. We demonstrated global heat kernel features to capture properties of network efficiency with the discriminative power to classify between different network topologies. In addition, we showed the efficacy of these features to predict motor dysfunction in a large cohort of preterm neonates. In summary, we have shown that energy transfer captured 'dynamically' by heat kernels may reveal aspects of network organisation which have the potential to serve as biomarkers for disease characterisation.

\section{Acknowledgements}

Data on preterm infants used in the paper was obtained as part of independent research commissioned by the National Institute for Health Research (NIHR). The views and opinions expressed by authors in this publication are those of the authors and do not necessarily reflect those of the NHS, the NIHR, MRIC, CCF, NETSCC, the Programme Grants for Applied Research programme or the Department of Health. The programme of research funded by the NIHR Programme Grants for Applied Research Programme (RP-PG-0707-10154.) and in the future a compendium report of the whole programme will be published in the NIHR Journal. The work was also supported by the NIHR Biomedical Research Centers at Guy’s and St Thomas’ NHS Trust and Imperial College Healthcare Trust.

\section*{References}

\bibliography{HK_ref_list2}

\begin{thebibliography}{10}
\expandafter\ifx\csname url\endcsname\relax
  \def\url#1{\texttt{#1}}\fi
\expandafter\ifx\csname urlprefix\endcsname\relax\def\urlprefix{URL }\fi
\expandafter\ifx\csname href\endcsname\relax
  \def\href#1#2{#2} \def\path#1{#1}\fi

\bibitem{bullmore_complex_2009}
E.~Bullmore, O.~Sporns, Complex brain networks: graph theoretical analysis of
  structural and functional systems, Nat Rev Neurosci 10~(3) (2009) 186--198.
\newblock \href {http://dx.doi.org/10.1038/nrn2575}
  {\path{doi:10.1038/nrn2575}}.

\bibitem{fornito_connectomics_2015}
A.~Fornito, A.~Zalesky, M.~Breakspear, The connectomics of brain disorders,
  Nature Reviews Neuroscience 16~(3) (2015) 159--172.
\newblock \href {http://dx.doi.org/10.1038/nrn3901}
  {\path{doi:10.1038/nrn3901}}.

\bibitem{jones_white_2013}
D.~K. Jones, T.~R. Knösche, R.~Turner, White matter integrity, fiber count,
  and other fallacies: the do's and don'ts of diffusion {MRI}, Neuroimage 73
  (2013) 239--254.
\newblock \href {http://dx.doi.org/10.1016/j.neuroimage.2012.06.081}
  {\path{doi:10.1016/j.neuroimage.2012.06.081}}.

\bibitem{fornito_graph_2013}
A.~Fornito, A.~Zalesky, M.~Breakspear, Graph analysis of the human connectome:
  {Promise}, progress, and pitfalls, NeuroImage 80 (2013) 426--444.
\newblock \href {http://dx.doi.org/10.1016/j.neuroimage.2013.04.087}
  {\path{doi:10.1016/j.neuroimage.2013.04.087}}.

\bibitem{van_den_heuvel_small-world_2008}
M.~P. van~den Heuvel, C.~J. Stam, M.~Boersma, H.~E. Hulshoff~Pol, Small-world
  and scale-free organization of voxel-based resting-state functional
  connectivity in the human brain, Neuroimage 43~(3) (2008) 528--539.
\newblock \href {http://dx.doi.org/10.1016/j.neuroimage.2008.08.010}
  {\path{doi:10.1016/j.neuroimage.2008.08.010}}.

\bibitem{towlson_rich_2013}
E.~K. Towlson, P.~E. Vértes, S.~E. Ahnert, W.~R. Schafer, E.~T. Bullmore, The
  rich club of the {C}. elegans neuronal connectome, J. Neurosci. 33~(15)
  (2013) 6380--6387.
\newblock \href {http://dx.doi.org/10.1523/JNEUROSCI.3784-12.2013}
  {\path{doi:10.1523/JNEUROSCI.3784-12.2013}}.

\bibitem{ball_rich-club_2014}
G.~Ball, P.~Aljabar, S.~Zebari, N.~Tusor, T.~Arichi, N.~Merchant, E.~C.
  Robinson, E.~Ogundipe, D.~Rueckert, A.~D. Edwards, S.~J. Counsell, Rich-club
  organization of the newborn human brain, PNAS 111~(20) (2014) 7456--7461.
\newblock \href {http://dx.doi.org/10.1073/pnas.1324118111}
  {\path{doi:10.1073/pnas.1324118111}}.

\bibitem{watts_collective_1998}
D.~J. Watts, S.~H. Strogatz, Collective dynamics of 'small-world' networks,
  Nature 393~(6684) (1998) 440--442.
\newblock \href {http://dx.doi.org/10.1038/30918} {\path{doi:10.1038/30918}}.

\bibitem{heuvel_high-cost_2012}
M.~P. v.~d. Heuvel, R.~S. Kahn, J.~Goñi, O.~Sporns, High-cost, high-capacity
  backbone for global brain communication, PNAS 109~(28) (2012) 11372--11377.
\newblock \href {http://dx.doi.org/10.1073/pnas.1203593109}
  {\path{doi:10.1073/pnas.1203593109}}.

\bibitem{colizza_detecting_2006}
V.~Colizza, A.~Flammini, M.~A. Serrano, A.~Vespignani, Detecting rich-club
  ordering in complex networks, Nat Phys 2~(2) (2006) 110--115.
\newblock \href {http://dx.doi.org/10.1038/nphys209}
  {\path{doi:10.1038/nphys209}}.

\bibitem{collin_structural_2014}
G.~Collin, O.~Sporns, R.~C.~W. Mandl, M.~P. van~den Heuvel, Structural and
  functional aspects relating to cost and benefit of rich club organization in
  the human cerebral cortex, Cereb. Cortex 24~(9) (2014) 2258--2267.
\newblock \href {http://dx.doi.org/10.1093/cercor/bht064}
  {\path{doi:10.1093/cercor/bht064}}.

\bibitem{mcauley_rich-club_2007}
J.~J. McAuley, L.~d.~F. Costa, T.~S. Caetano, Rich-club phenomenon across
  complex network hierarchies, Applied Physics Letters 91~(8) (2007) 084103.
\newblock \href {http://dx.doi.org/10.1063/1.2773951}
  {\path{doi:10.1063/1.2773951}}.

\bibitem{rubinov_complex_2010}
M.~Rubinov, O.~Sporns, Complex network measures of brain connectivity: {Uses}
  and interpretations, NeuroImage 52~(3) (2010) 1059--1069.
\newblock \href {http://dx.doi.org/10.1016/j.neuroimage.2009.10.003}
  {\path{doi:10.1016/j.neuroimage.2009.10.003}}.

\bibitem{saxena_selective_2011}
S.~Saxena, P.~Caroni, Selective neuronal vulnerability in neurodegenerative
  diseases: from stressor thresholds to degeneration, Neuron 71~(1) (2011)
  35--48.
\newblock \href {http://dx.doi.org/10.1016/j.neuron.2011.06.031}
  {\path{doi:10.1016/j.neuron.2011.06.031}}.

\bibitem{hirokawa_molecular_2010}
N.~Hirokawa, S.~Niwa, Y.~Tanaka, Molecular motors in neurons: transport
  mechanisms and roles in brain function, development, and disease, Neuron
  68~(4) (2010) 610--638.
\newblock \href {http://dx.doi.org/10.1016/j.neuron.2010.09.039}
  {\path{doi:10.1016/j.neuron.2010.09.039}}.

\bibitem{perlson_retrograde_2010}
E.~Perlson, S.~Maday, M.-M. Fu, A.~J. Moughamian, E.~L.~F. Holzbaur, Retrograde
  axonal transport: pathways to cell death?, Trends Neurosci. 33~(7) (2010)
  335--344.
\newblock \href {http://dx.doi.org/10.1016/j.tins.2010.03.006}
  {\path{doi:10.1016/j.tins.2010.03.006}}.

\bibitem{rehme_cerebral_2013}
A.~K. Rehme, C.~Grefkes, Cerebral network disorders after stroke: evidence from
  imaging-based connectivity analyses of active and resting brain states in
  humans, J. Physiol. (Lond.) 591~(Pt 1) (2013) 17--31.
\newblock \href {http://dx.doi.org/10.1113/jphysiol.2012.243469}
  {\path{doi:10.1113/jphysiol.2012.243469}}.

\bibitem{tabrizi_biological_2009}
S.~J. Tabrizi, D.~R. Langbehn, B.~R. Leavitt, R.~A. Roos, A.~Durr, D.~Craufurd,
  C.~Kennard, S.~L. Hicks, N.~C. Fox, R.~I. Scahill, B.~Borowsky, A.~J. Tobin,
  H.~D. Rosas, H.~Johnson, R.~Reilmann, B.~Landwehrmeyer, J.~C. Stout,
  {TRACK-HD investigators}, Biological and clinical manifestations of
  {Huntington}'s disease in the longitudinal {TRACK}-{HD} study:
  cross-sectional analysis of baseline data, Lancet Neurol 8~(9) (2009)
  791--801.
\newblock \href {http://dx.doi.org/10.1016/S1474-4422(09)70170-X}
  {\path{doi:10.1016/S1474-4422(09)70170-X}}.

\bibitem{goedert_100_2013}
M.~Goedert, M.~G. Spillantini, K.~Del~Tredici, H.~Braak, 100 years of {Lewy}
  pathology, Nat Rev Neurol 9~(1) (2013) 13--24.
\newblock \href {http://dx.doi.org/10.1038/nrneurol.2012.242}
  {\path{doi:10.1038/nrneurol.2012.242}}.

\bibitem{sleimen-malkoun_aging_2014}
R.~Sleimen-Malkoun, J.-J. Temprado, S.~L. Hong, Aging induced loss of
  complexity and dedifferentiation: consequences for coordination dynamics
  within and between brain, muscular and behavioral levels, Front Aging
  Neurosci 6.
\newblock \href {http://dx.doi.org/10.3389/fnagi.2014.00140}
  {\path{doi:10.3389/fnagi.2014.00140}}.

\bibitem{honey_functional_2005}
G.~D. Honey, E.~Pomarol-Clotet, P.~R. Corlett, R.~A.~E. Honey, P.~J. McKenna,
  E.~T. Bullmore, P.~C. Fletcher, Functional dysconnectivity in schizophrenia
  associated with attentional modulation of motor function, Brain 128~(Pt 11)
  (2005) 2597--2611.
\newblock \href {http://dx.doi.org/10.1093/brain/awh632}
  {\path{doi:10.1093/brain/awh632}}.

\bibitem{chiaravalloti_cognitive_2015}
N.~D. Chiaravalloti, H.~M. Genova, J.~DeLuca, Cognitive {Rehabilitation} in
  {Multiple} {Sclerosis}: {The} {Role} of {Plasticity}, Front Neurol 6.
\newblock \href {http://dx.doi.org/10.3389/fneur.2015.00067}
  {\path{doi:10.3389/fneur.2015.00067}}.

\bibitem{elman_neural_2014}
J.~A. Elman, H.~Oh, C.~M. Madison, S.~L. Baker, J.~W. Vogel, S.~M. Marks,
  S.~Crowley, J.~P. O'Neil, W.~J. Jagust, Neural compensation in older people
  with brain amyloid-β deposition, Nat Neurosci 17~(10) (2014) 1316--1318.
\newblock \href {http://dx.doi.org/10.1038/nn.3806}
  {\path{doi:10.1038/nn.3806}}.

\bibitem{schoonheim_network_2015}
M.~M. Schoonheim, K.~A. Meijer, J.~J.~G. Geurts, Network collapse and cognitive
  impairment in multiple sclerosis, Front. Neurol. 6 (2015) 82.
\newblock \href {http://dx.doi.org/10.3389/fneur.2015.00082}
  {\path{doi:10.3389/fneur.2015.00082}}.

\bibitem{odish_dynamics_2015}
O.~F. Odish, K.~Caeyenberghs, H.~Hosseini, S.~J. van~den Bogaard, R.~A. Roos,
  A.~Leemans, Dynamics of the connectome in {Huntington}'s disease: {A}
  longitudinal diffusion {MRI} study, Neuroimage Clin 9 (2015) 32--43.
\newblock \href {http://dx.doi.org/10.1016/j.nicl.2015.07.003}
  {\path{doi:10.1016/j.nicl.2015.07.003}}.

\bibitem{lo_diffusion_2010}
C.-Y. Lo, P.-N. Wang, K.-H. Chou, J.~Wang, Y.~He, C.-P. Lin, Diffusion tensor
  tractography reveals abnormal topological organization in structural cortical
  networks in {Alzheimer}'s disease, J. Neurosci. 30~(50) (2010) 16876--16885.
\newblock \href {http://dx.doi.org/10.1523/JNEUROSCI.4136-10.2010}
  {\path{doi:10.1523/JNEUROSCI.4136-10.2010}}.

\bibitem{wang_altered_2009}
L.~Wang, C.~Zhu, Y.~He, Y.~Zang, Q.~Cao, H.~Zhang, Q.~Zhong, Y.~Wang, Altered
  small-world brain functional networks in children with
  attention-deficit/hyperactivity disorder, Hum Brain Mapp 30~(2) (2009)
  638--649.
\newblock \href {http://dx.doi.org/10.1002/hbm.20530}
  {\path{doi:10.1002/hbm.20530}}.

\bibitem{pandit_whole-brain_2014}
A.~S. Pandit, E.~Robinson, P.~Aljabar, G.~Ball, I.~S. Gousias, Z.~Wang, J.~V.
  Hajnal, D.~Rueckert, S.~J. Counsell, G.~Montana, A.~D. Edwards, Whole-brain
  mapping of structural connectivity in infants reveals altered connection
  strength associated with growth and preterm birth, Cereb. Cortex 24~(9)
  (2014) 2324--2333.
\newblock \href {http://dx.doi.org/10.1093/cercor/bht086}
  {\path{doi:10.1093/cercor/bht086}}.

\bibitem{zalesky_network-based_2010}
A.~Zalesky, A.~Fornito, E.~T. Bullmore, Network-based statistic: identifying
  differences in brain networks, Neuroimage 53~(4) (2010) 1197--1207.
\newblock \href {http://dx.doi.org/10.1016/j.neuroimage.2010.06.041}
  {\path{doi:10.1016/j.neuroimage.2010.06.041}}.

\bibitem{shen_discriminative_2010}
H.~Shen, L.~Wang, Y.~Liu, D.~Hu, Discriminative analysis of resting-state
  functional connectivity patterns of schizophrenia using low dimensional
  embedding of {fMRI}, Neuroimage 49~(4) (2010) 3110--3121.
\newblock \href {http://dx.doi.org/10.1016/j.neuroimage.2009.11.011}
  {\path{doi:10.1016/j.neuroimage.2009.11.011}}.

\bibitem{arbabshirani_classification_2013}
M.~R. Arbabshirani, K.~A. Kiehl, G.~D. Pearlson, V.~D. Calhoun, Classification
  of schizophrenia patients based on resting-state functional network
  connectivity, Front Neurosci 7 (2013) 133.
\newblock \href {http://dx.doi.org/10.3389/fnins.2013.00133}
  {\path{doi:10.3389/fnins.2013.00133}}.

\bibitem{richiardi_classifying_2012}
J.~Richiardi, M.~Gschwind, S.~Simioni, J.-M. Annoni, B.~Greco, P.~Hagmann,
  M.~Schluep, P.~Vuilleumier, D.~Van De~Ville, Classifying minimally disabled
  multiple sclerosis patients from resting state functional connectivity,
  Neuroimage 62~(3) (2012) 2021--2033.
\newblock \href {http://dx.doi.org/10.1016/j.neuroimage.2012.05.078}
  {\path{doi:10.1016/j.neuroimage.2012.05.078}}.

\bibitem{prasad_brain_2015}
G.~Prasad, S.~H. Joshi, T.~M. Nir, A.~W. Toga, P.~M. Thompson, Brain
  connectivity and novel network measures for {Alzheimer}'s disease
  classification, Neurobiology of Aging 36 (2015) S121--S131.
\newblock \href {http://dx.doi.org/10.1016/j.neurobiolaging.2014.04.037}
  {\path{doi:10.1016/j.neurobiolaging.2014.04.037}}.

\bibitem{rosa_sparse_2015}
M.~J. Rosa, L.~Portugal, T.~Hahn, A.~J. Fallgatter, M.~I. Garrido,
  J.~Shawe-Taylor, J.~Mourao-Miranda, Sparse network-based models for patient
  classification using {fMRI}, Neuroimage 105 (2015) 493--506.
\newblock \href {http://dx.doi.org/10.1016/j.neuroimage.2014.11.021}
  {\path{doi:10.1016/j.neuroimage.2014.11.021}}.

\bibitem{sacchet_support_2015}
M.~D. Sacchet, G.~Prasad, L.~C. Foland-Ross, P.~M. Thompson, I.~H. Gotlib,
  Support vector machine classification of major depressive disorder using
  diffusion-weighted neuroimaging and graph theory, Front Psychiatry 6 (2015)
  21.
\newblock \href {http://dx.doi.org/10.3389/fpsyt.2015.00021}
  {\path{doi:10.3389/fpsyt.2015.00021}}.

\bibitem{meskaldji_improved_2015}
D.-E. Meskaldji, L.~Vasung, D.~Romascano, J.-P. Thiran, P.~Hagmann,
  S.~Morgenthaler, D.~Van De~Ville, Improved statistical evaluation of group
  differences in connectomes by screening-filtering strategy with application
  to study maturation of brain connections between childhood and adolescence,
  Neuroimage 108 (2015) 251--264.
\newblock \href {http://dx.doi.org/10.1016/j.neuroimage.2014.11.059}
  {\path{doi:10.1016/j.neuroimage.2014.11.059}}.

\bibitem{babaud_uniqueness_1986}
J.~Babaud, A.~P. Witkin, M.~Baudin, R.~O. Duda, Uniqueness of the gaussian
  kernel for scale-space filtering, IEEE Trans Pattern Anal Mach Intell 8~(1)
  (1986) 26--33.

\bibitem{perona_scale-space_1990}
P.~Perona, J.~Malik, Scale-space and edge detection using anisotropic
  diffusion, IEEE Transactions on Pattern Analysis and Machine Intelligence
  12~(7) (1990) 629--639.
\newblock \href {http://dx.doi.org/10.1109/34.56205}
  {\path{doi:10.1109/34.56205}}.

\bibitem{zhang_graph_2008}
F.~Zhang, E.~R. Hancock, Graph {Spectral} {Image} {Smoothing} {Using} the
  {Heat} {Kernel}, Pattern Recogn. 41~(11) (2008) 3328--3342.
\newblock \href {http://dx.doi.org/10.1016/j.patcog.2008.05.007}
  {\path{doi:10.1016/j.patcog.2008.05.007}}.

\bibitem{raj_network_2012}
A.~Raj, A.~Kuceyeski, M.~Weiner, A {Network} {Diffusion} {Model} of {Disease}
  {Progression} in {Dementia}, Neuron 73~(6) (2012) 1204--1215.
\newblock \href {http://dx.doi.org/10.1016/j.neuron.2011.12.040}
  {\path{doi:10.1016/j.neuron.2011.12.040}}.

\bibitem{abdelnour_network_2014}
F.~Abdelnour, H.~U. Voss, A.~Raj, Network diffusion accurately models the
  relationship between structural and functional brain connectivity networks,
  NeuroImage 90 (2014) 335--347.
\newblock \href {http://dx.doi.org/10.1016/j.neuroimage.2013.12.039}
  {\path{doi:10.1016/j.neuroimage.2013.12.039}}.

\bibitem{misic_cooperative_2015}
B.~Mišić, R.~F. Betzel, A.~Nematzadeh, J.~Goñi, A.~Griffa, P.~Hagmann,
  A.~Flammini, Y.-Y. Ahn, O.~Sporns, Cooperative and {Competitive} {Spreading}
  {Dynamics} on the {Human} {Connectome}, Neuron 86~(6) (2015) 1518--1529.
\newblock \href {http://dx.doi.org/10.1016/j.neuron.2015.05.035}
  {\path{doi:10.1016/j.neuron.2015.05.035}}.

\bibitem{delobel-ayoub_behavioral_2009}
M.~Delobel-Ayoub, C.~Arnaud, M.~White-Koning, C.~Casper, V.~Pierrat, M.~Garel,
  A.~Burguet, J.-C. Roze, J.~Matis, J.-C. Picaud, M.~Kaminski, B.~Larroque,
  {EPIPAGE Study Group}, Behavioral problems and cognitive performance at 5
  years of age after very preterm birth: the {EPIPAGE} {Study}, Pediatrics
  123~(6) (2009) 1485--1492.
\newblock \href {http://dx.doi.org/10.1542/peds.2008-1216}
  {\path{doi:10.1542/peds.2008-1216}}.

\bibitem{edwards_developmental_2011}
J.~Edwards, M.~Berube, K.~Erlandson, S.~Haug, H.~Johnstone, M.~Meagher,
  S.~Sarkodee-Adoo, J.~G. Zwicker, Developmental coordination disorder in
  school-aged children born very preterm and/or at very low birth weight: a
  systematic review, J Dev Behav Pediatr 32~(9) (2011) 678--687.
\newblock \href {http://dx.doi.org/10.1097/DBP.0b013e31822a396a}
  {\path{doi:10.1097/DBP.0b013e31822a396a}}.

\bibitem{marlow_motor_2007}
N.~Marlow, E.~M. Hennessy, M.~A. Bracewell, D.~Wolke, {EPICure Study Group},
  Motor and executive function at 6 years of age after extremely preterm birth,
  Pediatrics 120~(4) (2007) 793--804.
\newblock \href {http://dx.doi.org/10.1542/peds.2007-0440}
  {\path{doi:10.1542/peds.2007-0440}}.

\bibitem{back_brain_2014}
S.~A. Back, S.~P. Miller, Brain injury in premature neonates: {A} primary
  cerebral dysmaturation disorder?, Ann. Neurol. 75~(4) (2014) 469--486.
\newblock \href {http://dx.doi.org/10.1002/ana.24132}
  {\path{doi:10.1002/ana.24132}}.

\bibitem{van_kooij_neonatal_2012}
B.~J.~M. van Kooij, L.~S. de~Vries, G.~Ball, I.~C. van Haastert, M.~J. N.~L.
  Benders, F.~Groenendaal, S.~J. Counsell, Neonatal tract-based spatial
  statistics findings and outcome in preterm infants, AJNR Am J Neuroradiol
  33~(1) (2012) 188--194.
\newblock \href {http://dx.doi.org/10.3174/ajnr.A2723}
  {\path{doi:10.3174/ajnr.A2723}}.

\bibitem{duerden_tract-based_2015}
E.~G. Duerden, J.~Foong, V.~Chau, H.~Branson, K.~J. Poskitt, R.~E. Grunau,
  A.~Synnes, J.~G. Zwicker, S.~P. Miller, Tract-{Based} {Spatial} {Statistics}
  in {Preterm}-{Born} {Neonates} {Predicts} {Cognitive} and {Motor} {Outcomes}
  at 18 {Months}, AJNR Am J Neuroradiol 36~(8) (2015) 1565--1571.
\newblock \href {http://dx.doi.org/10.3174/ajnr.A4312}
  {\path{doi:10.3174/ajnr.A4312}}.

\bibitem{ball_thalamocortical_2015}
G.~Ball, L.~Pazderova, A.~Chew, N.~Tusor, N.~Merchant, T.~Arichi, J.~M. Allsop,
  F.~M. Cowan, A.~D. Edwards, S.~J. Counsell, Thalamocortical {Connectivity}
  {Predicts} {Cognition} in {Children} {Born} {Preterm}, Cereb. Cortex 25~(11)
  (2015) 4310--4318.
\newblock \href {http://dx.doi.org/10.1093/cercor/bhu331}
  {\path{doi:10.1093/cercor/bhu331}}.

\bibitem{chung_spectral_1997}
F.~R.~K. Chung, Spectral {Graph} {Theory}, CBMS, Number 92, American
  Mathematical Society, 1997.

\bibitem{yau_lectures_1994}
S.~T. Yau, R.~M. Schoen, Lectures on {Differential} {Geometry}, International
  Press Inc., Boston, 1994.

\bibitem{fiedler_laplacian_1989}
M.~Fiedler, Laplacian of graphs and algebraic connectivity, Combinatorics and
  Graph Theory 25~(1) (1989) 57--70.

\bibitem{al-mohy_new_2009}
A.~Al-Mohy, N.~Higham, A {New} {Scaling} and {Squaring} {Algorithm} for the
  {Matrix} {Exponential}, SIAM. J. Matrix Anal. \& Appl. 31~(3) (2009)
  970--989.
\newblock \href {http://dx.doi.org/10.1137/09074721X}
  {\path{doi:10.1137/09074721X}}.

\bibitem{erdos_random_1959}
P.~Erdős, A.~Rényi, On random graphs, Publicationes Mathematicae 6 (1959)
  290--297.

\bibitem{bridson_fast_2007}
R.~Bridson, Fast {Poisson} disk sampling in arbitrary dimensions, in: {ACM}
  {SIGGRAPH}, 2007.

\bibitem{schirmer_parcellation-independent_2014}
M.~D. Schirmer, G.~Ball, S.~J. Counsell, A.~D. Edwards, D.~Rueckert, J.~V.
  Hajnal, P.~Aljabar, Parcellation-{Independent} {Multi}-{Scale} {Framework}
  for {Brain} {Network} {Analysis}, in: MICCAI{Computational} {Diffusion}
  {MRI}, Mathematics and {Visualization}, Springer International Publishing,
  2014, pp. 23--32, dOI: 10.1007/978-3-319-11182-7\_3.

\bibitem{simard_fastest_2005}
D.~Simard, L.~Nadeau, H.~Kröger, Fastest learning in small-world neural
  networks, Physics Letters A 336~(1) (2005) 8--15.
\newblock \href {http://dx.doi.org/10.1016/j.physleta.2004.12.078}
  {\path{doi:10.1016/j.physleta.2004.12.078}}.

\bibitem{bassett_adaptive_2006}
D.~S. Bassett, A.~Meyer-Lindenberg, S.~Achard, T.~Duke, E.~Bullmore, Adaptive
  reconfiguration of fractal small-world human brain functional networks, Proc.
  Natl. Acad. Sci. U.S.A. 103~(51) (2006) 19518--19523.
\newblock \href {http://dx.doi.org/10.1073/pnas.0606005103}
  {\path{doi:10.1073/pnas.0606005103}}.

\bibitem{savitzky_smoothing_1964}
A.~Savitzky, M.~J.~E. Golay, Smoothing and {Differentiation} of {Data} by
  {Simplified} {Least} {Squares} {Procedures}., Anal. Chem. 36~(8) (1964)
  1627--1639.
\newblock \href {http://dx.doi.org/10.1021/ac60214a047}
  {\path{doi:10.1021/ac60214a047}}.

\bibitem{pedregosa_scikit-learn:_2011}
F.~Pedregosa, G.~Varoquaux, A.~Gramfort, V.~Michel, B.~Thirion, O.~Grisel,
  M.~Blondel, P.~Prettenhofer, R.~Weiss, V.~Dubourg, J.~Vanderplas, A.~Passos,
  D.~Cournapeau, M.~Brucher, M.~Perrot, Ã.~Duchesnay, Scikit-learn: {Machine}
  {Learning} in {Python}, Journal of Machine Learning Research 12 (2011)
  2825−2830.

\bibitem{bayley_bayley_2006}
N.~Bayley, Bayley {Scales} of {Infant} and {Toddler} {Development}, {Harcourt},
  {San} {Antonio}, 3rd edition.

\bibitem{tzourio-mazoyer_automated_2002}
N.~Tzourio-Mazoyer, B.~Landeau, D.~Papathanassiou, F.~Crivello, O.~Etard,
  N.~Delcroix, B.~Mazoyer, M.~Joliot, Automated anatomical labeling of
  activations in {SPM} using a macroscopic anatomical parcellation of the {MNI}
  {MRI} single-subject brain, Neuroimage 15~(1) (2002) 273--289.
\newblock \href {http://dx.doi.org/10.1006/nimg.2001.0978}
  {\path{doi:10.1006/nimg.2001.0978}}.

\bibitem{behrens_probabilistic_2007}
T.~E.~J. Behrens, H.~J. Berg, S.~Jbabdi, M.~F.~S. Rushworth, M.~W. Woolrich,
  Probabilistic diffusion tractography with multiple fibre orientations: {What}
  can we gain?, Neuroimage 34~(1) (2007) 144--155.
\newblock \href {http://dx.doi.org/10.1016/j.neuroimage.2006.09.018}
  {\path{doi:10.1016/j.neuroimage.2006.09.018}}.

\bibitem{robinson_identifying_2010}
E.~C. Robinson, A.~Hammers, A.~Ericsson, A.~D. Edwards, D.~Rueckert,
  Identifying population differences in whole-brain structural networks: a
  machine learning approach, Neuroimage 50~(3) (2010) 910--919.
\newblock \href {http://dx.doi.org/10.1016/j.neuroimage.2010.01.019}
  {\path{doi:10.1016/j.neuroimage.2010.01.019}}.

\bibitem{catani_rises_2005}
M.~Catani, D.~H. Ffytche, The rises and falls of disconnection syndromes, Brain
  128~(10) (2005) 2224--2239.
\newblock \href {http://dx.doi.org/10.1093/brain/awh622}
  {\path{doi:10.1093/brain/awh622}}.

\bibitem{buckner_organization_2011}
R.~L. Buckner, F.~M. Krienen, A.~Castellanos, J.~C. Diaz, B.~T.~T. Yeo, The
  organization of the human cerebellum estimated by intrinsic functional
  connectivity, J. Neurophysiol. 106~(5) (2011) 2322--2345.
\newblock \href {http://dx.doi.org/10.1152/jn.00339.2011}
  {\path{doi:10.1152/jn.00339.2011}}.

\bibitem{vincent_intrinsic_2007}
J.~L. Vincent, G.~H. Patel, M.~D. Fox, A.~Z. Snyder, J.~T. Baker, D.~C.
  Van~Essen, J.~M. Zempel, L.~H. Snyder, M.~Corbetta, M.~E. Raichle, Intrinsic
  functional architecture in the anaesthetized monkey brain, Nature 447~(7140)
  (2007) 83--86.
\newblock \href {http://dx.doi.org/10.1038/nature05758}
  {\path{doi:10.1038/nature05758}}.

\bibitem{bullmore_brain_2011}
E.~T. Bullmore, D.~S. Bassett, Brain graphs: graphical models of the human
  brain connectome, Annu Rev Clin Psychol 7 (2011) 113--140.
\newblock \href {http://dx.doi.org/10.1146/annurev-clinpsy-040510-143934}
  {\path{doi:10.1146/annurev-clinpsy-040510-143934}}.

\bibitem{chau_abnormal_2013}
V.~Chau, A.~Synnes, R.~E. Grunau, K.~J. Poskitt, R.~Brant, S.~P. Miller,
  Abnormal brain maturation in preterm neonates associated with adverse
  developmental outcomes, Neurology 81~(24) (2013) 2082--2089.
\newblock \href {http://dx.doi.org/10.1212/01.wnl.0000437298.43688.b9}
  {\path{doi:10.1212/01.wnl.0000437298.43688.b9}}.

\bibitem{brown_prediction_2015}
C.~J. Brown, S.~P. Miller, B.~G. Booth, K.~J. Poskitt, V.~Chau, A.~R. Synnes,
  J.~G. Zwicker, R.~E. Grunau, G.~Hamarneh, Prediction of {Motor} {Function} in
  {Very} {Preterm} {Infants} {Using} {Connectome} {Features} and {Local}
  {Synthetic} {Instances}, in: {MICCAI} 2015, no. 9349 in Lecture {Notes} in
  {Computer} {Science}, Springer International Publishing, 2015, pp. 69--76,
  dOI: 10.1007/978-3-319-24553-9\_9.

\bibitem{fornito_competitive_2012}
A.~Fornito, B.~J. Harrison, A.~Zalesky, J.~S. Simons, Competitive and
  cooperative dynamics of large-scale brain functional networks supporting
  recollection, PNAS 109~(31) (2012) 12788--12793.
\newblock \href {http://dx.doi.org/10.1073/pnas.1204185109}
  {\path{doi:10.1073/pnas.1204185109}}.

\bibitem{hellyer_control_2014}
P.~J. Hellyer, M.~Shanahan, G.~Scott, R.~J.~S. Wise, D.~J. Sharp, R.~Leech, The
  {Control} of {Global} {Brain} {Dynamics}: {Opposing} {Actions} of
  {Frontoparietal} {Control} and {Default} {Mode} {Networks} on {Attention}, J.
  Neurosci. 34~(2) (2014) 451--461.
\newblock \href {http://dx.doi.org/10.1523/JNEUROSCI.1853-13.2014}
  {\path{doi:10.1523/JNEUROSCI.1853-13.2014}}.

\bibitem{kragel_functional_2015}
J.~E. Kragel, S.~M. Polyn, Functional interactions between large-scale networks
  during memory search, Cereb. Cortex 25~(3) (2015) 667--679.
\newblock \href {http://dx.doi.org/10.1093/cercor/bht258}
  {\path{doi:10.1093/cercor/bht258}}.

\end{thebibliography}

\clearpage
\section{Appendix}

\subsection{Classification Pipelines}
\label{app:Algs}

\begin{algorithm}
\caption{Synthetic networks classification pipeline}
\begin{algorithmic}[1]
\For{each density $d_{sw} = [15, 20, 30, 40\%]$}
\State Calculate candidate $sw(d_{sw},p_{sw})$ topology where $p_{sw} = p:min\abs{t_{c}^g}_{p_{1}}^{p_{100}}$
\For{each remaining topologies, $(d,p)$}
\For{each partition, $[edge, \neg edge]$}
\For{each feature set}
\State Classify $sw(d_{sw},p_{sw})$ vs $(d,p)$ with 10-fold cross-validation
\State Store mean classification performance scores.
\EndFor
\EndFor
\EndFor
\EndFor
\end{algorithmic}
\end{algorithm}
\label{app:alg_synthetic}

\begin{algorithm}
\caption{ePrime classification pipeline: $[HK, GA, SA]$ denote classifying with heat kernel features, heat kernel features with GA regressed from the data, and with SA regressed from the data.}
\begin{algorithmic}[1]

\For{each experiment, $[HK, GA, SA]$}
\For{each repetition, $r=0:5$}
\State Define folds for 10-fold cross-validation
\For{each partition, $[edge, \neg edge]$}
\For{each feature set}
\State Classify \textit{low motor function} vs \textit{controls}
\State Store mean classification performance scores.
\EndFor
\EndFor
\EndFor
\EndFor
\State Compute average classification performance scores across all $r$
\end{algorithmic}
\end{algorithm}
\label{app:alg_ePrime}
\clearpage
%% ================================================================================

\subsection{Standard network feature sets used for classifying ePrime data by motor ability}\label{app:standard_featuresets}

\begin{table}[h!]
	\begin{center}
		\caption{List of standard feature sets used for classification. CPL = characteristic path length, eBC = Average edge betweenness centrality, CC = Average clustering coefficient, $G_{eff}$ = Global efficiency}
		\label{tab:features_list}
		\begin{tabular}{l}
			\toprule
			\hline			
			Standard network feature sets\\
			\midrule
			1) Average CPL, eBC, CC, $G_{eff}$ \\	
			2) $Edge$ CPL, eBC, CC, $G_{eff}$  \\
			3) $\neg Edge$ CPL, eBC, CC, $G_{eff}$ \\
			4) $Edge$ CPL, $\neg Edge$ CPL, eBC, CC, $G_{eff}$ \\
			\bottomrule
		\end{tabular}
	\end{center}
\end{table}
\clearpage
%% ================================================================================

\subsection{Synthetic networks: Change in heat kernel values in $edge$ partitions}\label{app:HK_edge}

\begin{figure}[h!]
\centering
\includegraphics[width=1 \textwidth]{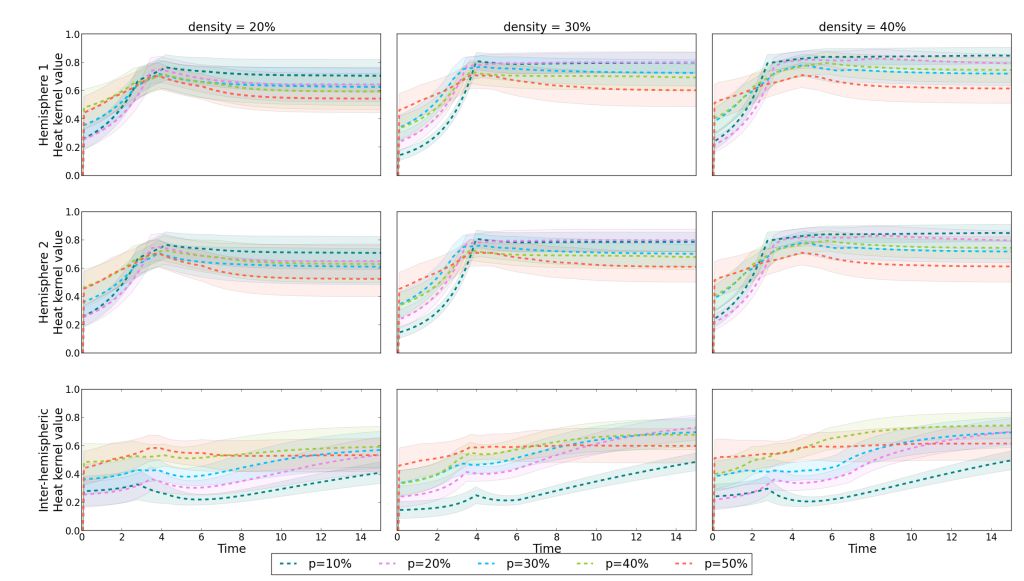}
\caption{Change in average heat kernel values for each $edge$ partition with time of a single synthetic network with densities $d$ = [20, 30, 40\%] at a range of randomisation percentages, $p$. Mean and standard deviation are over all heat kernel edges within each partition.}
\label{Afig:surr_HKwTime_edge}
\end{figure}

Figure~\ref{Afig:surr_HKwTime_edge} illustrates the change in heat for example surrogate networks with the time parameter, $t$. Mean heat kernel values, $H(t)_{u,v}$, in $edge$ partitions are plotted for networks with density $d$ =[20, 30, 40\%] and for a range of randomisation percentages.

\clearpage
%% ================================================================================

\subsection{Synthetic networks: Topology maps for each global $edge$ partition heat kernel feature}\label{app:HK_globalChange_edge}

\begin{figure}[h!]
\centering
\includegraphics[width=1 \textwidth]{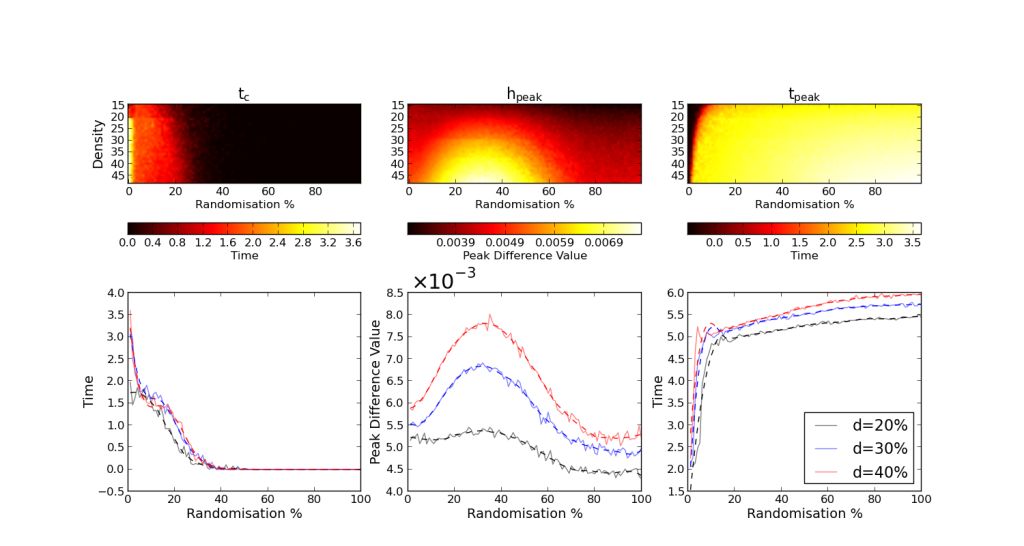}
\caption{Top row) Topology maps for each global heat kernel feature in the $edge$ partition. Each pixel in the map represents the mean (taken across all fifty replicates) heat kernel feature for a synthetic network topology with density, $d$ ($y$-axis), and randomisation percentage $p$ ($x$-axis). Bottom row) Example densities (i.e. rows) from the above corresponding topology maps depicting change in heat kernel feature with increasing randomisation percentage. Dashed line plots are interpolated global features across $p$ using a Savitzky-Golay filter.}
\label{Afig:surr_globalChange_edge}
\end{figure}

The top row in Figure~\ref{Afig:surr_globalChange_edge} contains topology maps showing global heat kernel features from $edge$ partitions across synthetic network replicates for all $(d,p)$. That is, each pixel in the map represents a $(d,p)$ topology, and contains the mean, global heat kernel feature over fifty replicates of $(d,p)$. The bottom row of Figure~\ref{Afig:surr_globalChange_edge} plots example rows in the corresponding topology map above, showing the change in mean heat kernel feature with randomisation for three example densities.

\clearpage
%% ================================================================================

\subsection{Synthetic networks: Classifier performance measures for candidate small-world networks versus all remaining topologies in the $\neg edge$ partition}\label{app:sw_class}

\begin{figure}[h!]
\centering
\includegraphics[width=1 \textwidth]{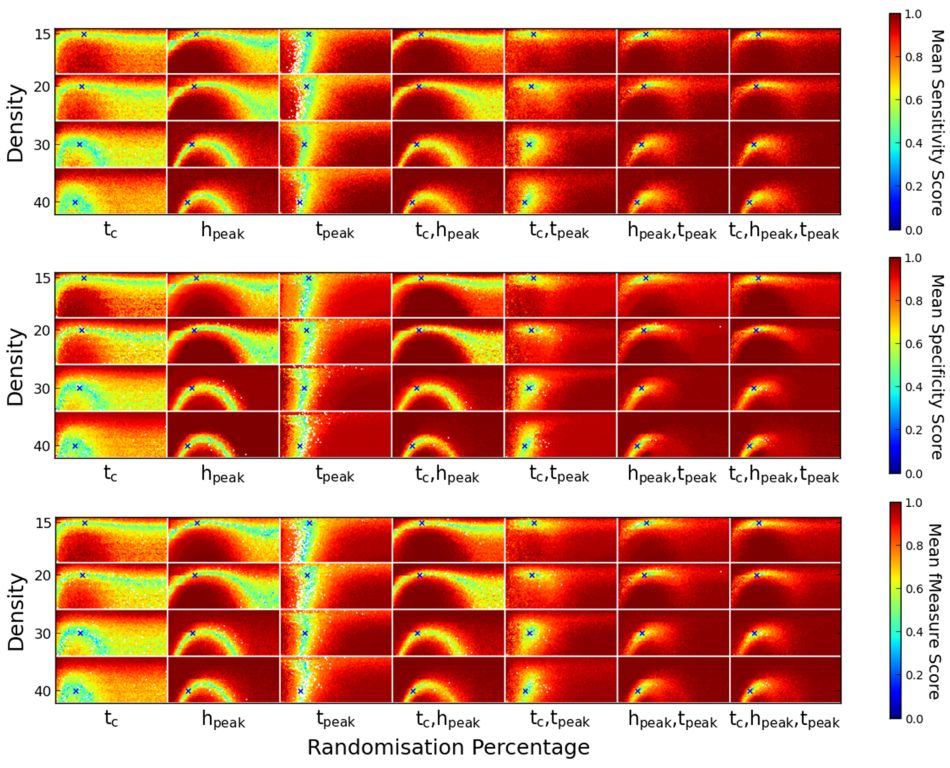}
\caption{LDA classifier performance maps - Results from classifying between sw(d,p) and all remaining topologies. From top to bottom, mean Sensitivity, Specificity and F-score results are presented for classifying by each of the seven feature sets calculated from the $\neg edge$ partition. The $sw(d,p)$ topologies computed for densities $d=[15,20,30,40\%]$ are presented in each row within a map, denoted by 'x'.}
\label{Afig:compiled_surr_class}
\end{figure}

\begin{figure}[h!]
\centering
\includegraphics[width=1 \textwidth]{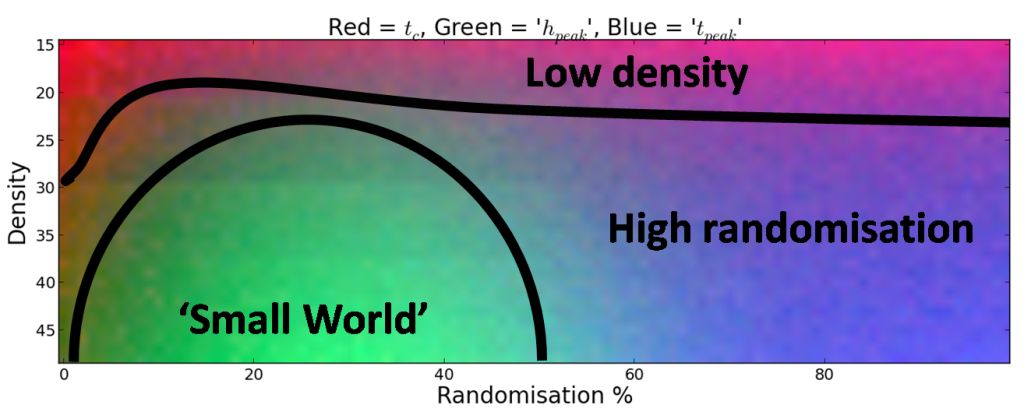}
\caption{RGB map delineating three broad regions across all synthetic networks generated which share similar topological features. $t_c$ is the red channel, $h_{peak}$ is the green channel and $t_{peak}$ the blue channel.}
\label{Afig:example_topological_regions}
\end{figure}

Figure~\ref{Afig:compiled_surr_class} plot classification performance measures of sensitivity, specificity and F-score from classifying global $\neg edge$ features between fifty candidate small-world topology surrogates against all remaining network topologies. Figure~\ref{Afig:example_topological_regions} is a representation of $t_c$, $h_{peak}$ and $t_{peak}$ combined, depicting three general regions across all generated synthetic networks which share similar topologies as captured by the heat kernel features.

\clearpage

%% ================================================================================

\subsection{Synthetic networks: Classifier performance measures for candidate random networks versus all remaining topologies in the $\neg edge$ partition}\label{app:rand_class}

\begin{figure}[h!]
\centering
\includegraphics[width=1 \textwidth]{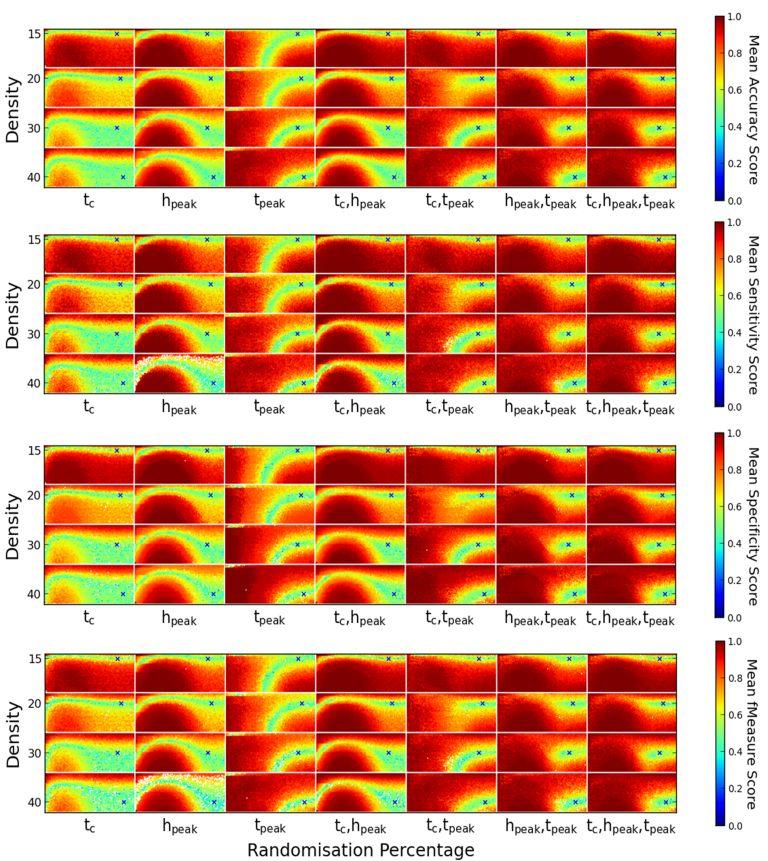}
\caption{LDA classifier performance maps - Results from classifying between $random(d,p_{rand}$) and all remaining topologies. From top to bottom, mean Accuracy, Sensitivity, Specificity and F-score results are presented for classifying by each of the seven feature sets calculated from the $\neg edge$ partition. The $random(d,p_{rand})$ topologies computed for densities $d=[15,20,30,40\%]$ are presented in each row within a map, denoted by 'x'.}
\label{Afig:compiled_surr_INEFFclass}
\end{figure}

The random topology, $random(d,p_{rand})$, was classified with every other surrogate topology. $random(d,p_{rand})$ was identified for $d = (15, 20, 30, 40\%)$ with $p_{rand} = p:\max\abs{t_{eq}}_{p=60\%}^{p=100\%}$ i.e. the peak of the $t_{c}$ curve following the 'small-world' dip in Figure~\ref{fig:surr_gridmaps} in the main article. We deem $random(d,p_{rand})$ to represent the network topology with density $d$ to be a candidate network which is inefficient for information transport. Figure~\ref{Afig:compiled_surr_INEFFclass} contains performance measures from classifying $random(d,p_{rand})$ against all remaining topologies.

\clearpage

%% ================================================================================

\subsection{Synthetic networks: Classifier performance measures for candidate small-world networks versus all remaining topologies in the $edge$ partition}\label{app:edge_sw_class}

\begin{figure}[h!]
\centering
\includegraphics[width=0.8 \textwidth]{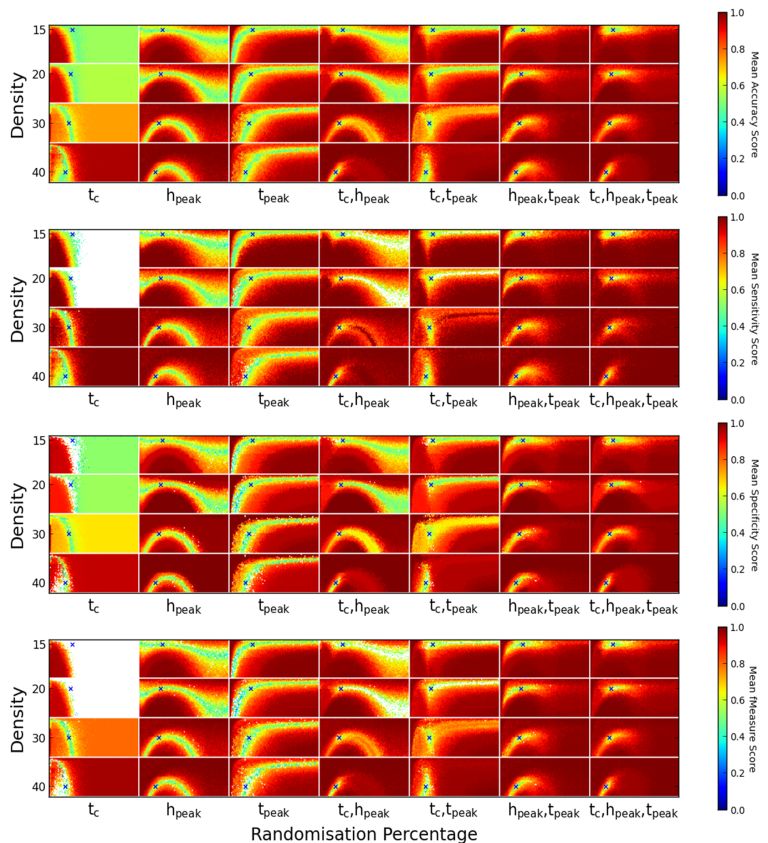}
\caption{LDA classifier performance maps - Results from classifying between $sw(d,p)$ and all remaining topologies. From top to bottom, mean Accuracy, Sensitivity, Specificity and F-score results are presented for classifying by each of the seven feature sets calculated from the $edge$ partition. The $sw(d,p)$ topologies computed for densities $d=[15,20,30,40\%]$ are presented in each row within a map, denoted by 'x'.}
\label{Afig:edge_surr_class_LDA}
\end{figure}

Figure~\ref{Afig:edge_surr_class_LDA} plot classification performance measures of sensitivity, specificity and F-score from classifying global $edge$ features between fifty candidate small-world topology surrogates (the same as those used in Figure~\ref{app:sw_class}) against all remaining network topologies. 

\clearpage
%% ================================================================================

\subsection{ePrime cohort: GNB classification performance on $\neg edge$ partitions}\label{app:ePrime_class_non-edge}

\begin{figure}[h!]
\centering
\includegraphics[width=0.7 \textwidth]{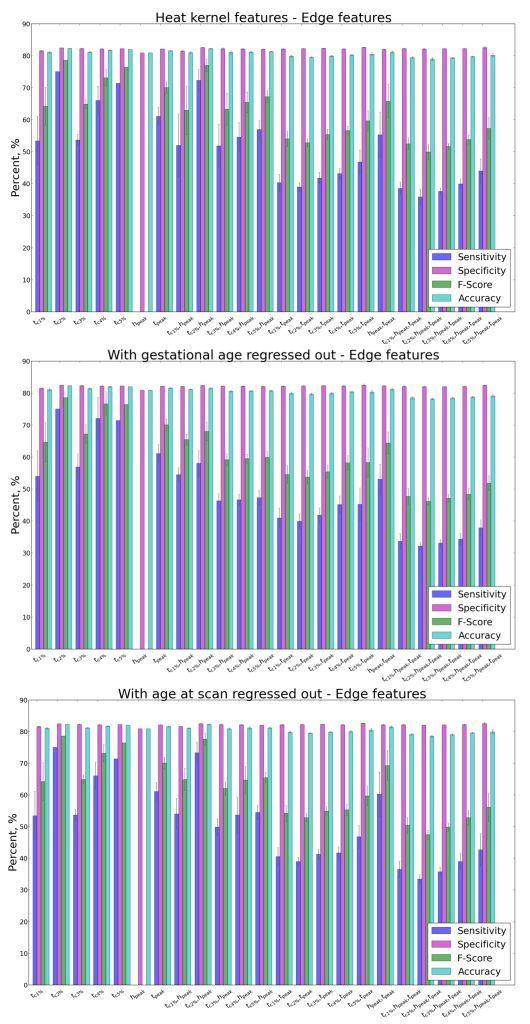}
\caption{GNB classification performance by feature set using heat kernel features for the $edge$ partition (top row). Each bar represents the average over five iterations of stratified 10-fold cross-validation. Error bars indicate standard deviation across iterations. Middle row are results from classification repeated on feature sets with GA regressed out. Bottom row contain results from classification repeated on feature sets with SA regressed out.}
\label{Afig:ePrime_fullFeatureSets_edge}
\end{figure}

\begin{figure}[h!]
\centering
\includegraphics[width=0.7 \textwidth]{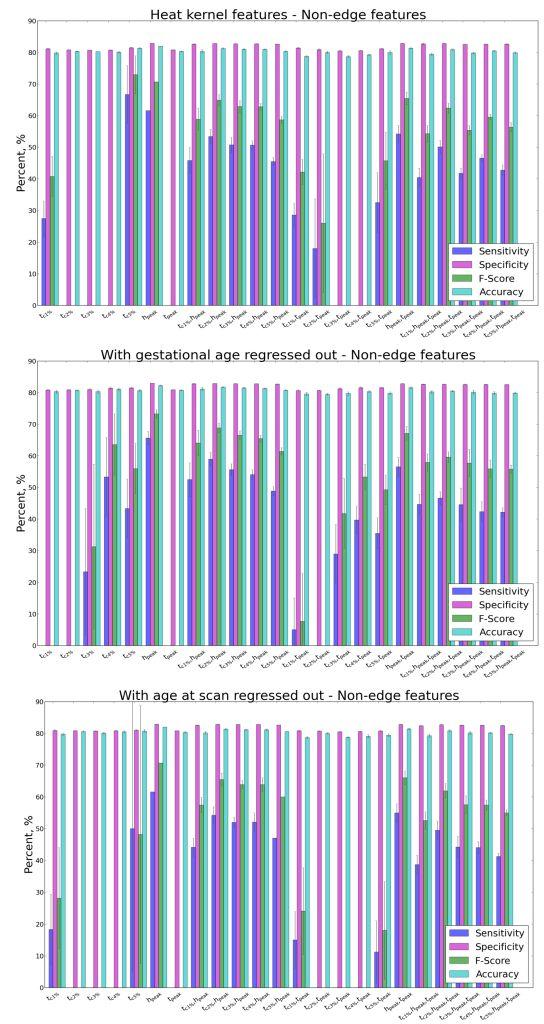}
\caption{GNB classification performance by feature set using heat kernel features for the $\neg edge$ partition (top row). Each bar represents the average over five iterations of stratified 10-fold cross-validation. Error bars indicate standard deviation across iterations. Middle row are results from classification repeated on feature sets with GA regressed out. Bottom row contain results from classification repeated on feature sets with SA regressed out.}
\label{Afig:ePrime_fullFeatureSets_nonEdge}
\end{figure}

GNB was performed on heat kernel features computed from structural networks to classify preterm infants by poor or normal motor function. Figures~\ref{Afig:ePrime_fullFeatureSets_edge} and ~\ref{Afig:ePrime_fullFeatureSets_nonEdge} show classification results for all feature sets tested as calculated from $edge$ and $\neg edge$ partitions, respectively. In addition, results for classification with age at scan and gestational age each separately regressed from the feature sets are presented.

\clearpage

%% ================================================================================

\subsection{ePrime cohort: GNB classification performance results}\label{app:ePrime_class_tables}

\begin{table}[h!]
	\begin{center}
		\caption{Classification scores based on heat kernel features in the $edge$ partition, averaged across five iterations of stratified 10-fold cross validation with GNB. * denote when all performance measures exceed 60\%.}
		\label{Atab:edge_HK_as_is}
		\begin{tabular}{lllll}
			\toprule
			\hline			
			Feature Set & Sensitivity & Specificity & F-Score & Accuracy\\
			\midrule
			$t_{c1\%}$ & 0.53429 & 0.81542 & 0.64233 & 0.81042\\
			$t_{c2\%}$ & 0.75 & 0.825 & 0.78571	0.82292*\\
			$t_{c3\%}$ & 0.53636 & 0.82251 & 0.64914 & 0.81181\\
			$t_{c4\%}$ & 0.66071 & 0.82168 & 0.73168 & 0.81736*\\
			$t_{c5\%}$ & 0.71429 & 0.82206 & 0.76439 & 0.81944*\\
			$h_{peak}$ & 0.0 & 0.80903 & 0.0 & 0.80903\\
			$t_{peak}$ & 0.61111 & 0.8213 & 0.70042 & 0.81528\\
			$t_{c1\%}$ and $h_{peak}$ & 0.52 & 0.81484 & 0.62942 & 0.80972\\
			$t_{c2\%}$ and $h_{peak}$ & 0.72333 & 0.82534 & 0.77054 & 0.82222*\\
			$t_{c3\%}$ and $h_{peak}$ & 0.51818 & 0.82225 & 0.63334 & 0.81042\\
			$t_{c4\%}$ and $h_{peak}$ & 0.5452 & 0.82112 & 0.65422 & 0.81181\\
			$t_{c5\%}$ and $h_{peak}$ & 0.56944 & 0.82092 & 0.67207 & 0.81319\\
			$t_{c1\%}$ and $t_{peak}$ & 0.40321 & 0.82098 & 0.54034 & 0.79861\\
			$t_{c2\%}$ and $t_{peak}$ & 0.38937 & 0.82222 & 0.52834 & 0.79514\\
			$t_{c3\%}$ and $t_{peak}$ & 0.41748 & 0.82301 & 0.55373 & 0.79931\\
			$t_{c4\%}$ and $t_{peak}$ & 0.4318 & 0.82164 & 0.56592 & 0.80208\\
			$t_{c5\%}$ and $t_{peak}$ & 0.46824 & 0.82644 & 0.59694 & 0.80486\\
			$h_{peak}$ and $t_{peak}$ & 0.55317 & 0.81974 & 0.65795 & 0.81181\\
			$t_{c1\%}$ and $h_{peak}$ and $t_{peak}$ & 0.38569 & 0.82209 & 0.52475 & 0.79444\\
			$t_{c1\%}$ and $h_{peak}$ and $t_{peak}$ & 0.35889 & 0.82116 & 0.49898 & 0.78958\\
			$t_{c1\%}$ and $h_{peak}$ and $t_{peak}$ & 0.37661 & 0.82183 & 0.51644 & 0.79306\\
			$t_{c1\%}$ and $h_{peak}$ and $t_{peak}$ & 0.39984 & 0.82214 & 0.53785 & 0.79722\\
			$t_{c1\%}$ and $h_{peak}$ and $t_{peak}$ & 0.43955 & 0.8258 & 0.57276 & 0.80139\\
			\bottomrule
		\end{tabular}
	\end{center}
\end{table}

\begin{table}[h!]
	\begin{center}
		\caption{Classification scores based on heat kernel features with GA regressed out in the $edge$ partition, averaged across five iterations of stratified 10-fold cross validation with GNB. * denote when all performance measures exceed 60\%.}
		\label{Atab:edge_HK_GA}
		\begin{tabular}{lllll}
			\toprule
			\hline			
			Feature Set & Sensitivity & Specificity & F-Score & Accuracy\\
			\midrule
			$t_{c1\%}$ & 	0.54	&	0.81497	&	0.64606	&	0.81042	\\
			$t_{c2\%}$ & 	0.75	&	0.825	&	0.78571	&	0.82292	*\\
			$t_{c3\%}$ & 	0.56909	&	0.82289	&	0.67205	&	0.81389	\\
			$t_{c4\%}$ & 	0.72024	&	0.82206	&	0.76621	&	0.81944	*\\
			$t_{c5\%}$ & 	0.71429	&	0.82206	&	0.76439	&	0.81944	*\\
			$h_{peak}$ & 	0.0	&	0.80889	&	0.0	&	0.80833	\\
			$t_{peak}$ & 	0.61111	&	0.8213	&	0.70042	&	0.81528	*\\
			$t_{c1\%}$ and $h_{peak}$ & 	0.54444	&	0.8202	&	0.65421	&	0.81181	\\
			$t_{c2\%}$ and $h_{peak}$ & 	0.58	&	0.82349	&	0.67982	&	0.81458	\\
			$t_{c3\%}$ and $h_{peak}$ &	0.46264	&	0.82182	&	0.59169	&	0.80556	\\
			$t_{c4\%}$ and $h_{peak}$ & 	0.46643	&	0.82102	&	0.59473	&	0.80625	\\
			$t_{c5\%}$ and $h_{peak}$ & 	0.47273	&	0.82022	&	0.59948	&	0.80694	\\
			$t_{c1\%}$ and $t_{peak}$ & 	0.40952	&	0.82111	&	0.5458	&	0.79931	\\
			$t_{c2\%}$ and $t_{peak}$ & 	0.39909	&	0.82248	&	0.53702	&	0.79653	\\
			$t_{c3\%}$ and $t_{peak}$ & 	0.41806	&	0.82301	&	0.55407	&	0.79931	\\
			$t_{c4\%}$ and $t_{peak}$ & 	0.45124	&	0.82203	&	0.58217	&	0.80417	\\
			$t_{c5\%}$ and $t_{peak}$ & 	0.45216	&	0.82474	&	0.5824	&	0.80347	\\
			$h_{peak}$ and $t_{peak}$ & 	0.53049	&	0.82238	&	0.64378	&	0.81111	\\
			$t_{c1\%}$ and $h_{peak}$ and $t_{peak}$ & 	0.33645	&	0.82071	&	0.47677	&	0.78472	\\
			$t_{c1\%}$ and $h_{peak}$ and $t_{peak}$ & 	0.32148	&	0.81968	&	0.46172	&	0.78194	\\
			$t_{c1\%}$ and $h_{peak}$ and $t_{peak}$ & 	0.33061	&	0.82009	&	0.47112	&	0.78403	\\
			$t_{c1\%}$ and $h_{peak}$ and $t_{peak}$ & 	0.34351	&	0.82028	&	0.48398	&	0.7875	\\
			$t_{c1\%}$ and $h_{peak}$ and $t_{peak}$ & 	0.37888	&	0.82383	&	0.51861	&	0.79097	\\
			\bottomrule
		\end{tabular}
	\end{center}
\end{table}

\begin{table}[h!]
	\begin{center}
		\caption{Classification scores based on heat kernel features with SA regressed out in the $edge$ partition, averaged across five iterations of stratified 10-fold cross validation with GNB. * denote when all performance measures exceed 60\%.}
		\label{Atab:edge_HK_SA}
		\begin{tabular}{lllll}
			\toprule
			\hline			
			Feature Set & Sensitivity & Specificity & F-Score & Accuracy\\
			\midrule
			$t_{c1\%}$ & 	0.53429	&	0.81542	&	0.64233	&	0.81042	\\
			$t_{c2\%}$ & 	0.75	&	0.825	&	0.78571	&	0.82292	*\\
			$t_{c3\%}$ & 	0.53636	&	0.82251	&	0.64914	&	0.81181	\\
			$t_{c4\%}$ & 	0.66071	&	0.82168	&	0.73168	&	0.81736	*\\
			$t_{c5\%}$ & 	0.71429	&	0.82206	&	0.76439	&	0.81944	*\\
			$h_{peak}$ & 	0.0	&	0.80903	&	0.0	&	0.80903	\\
			$t_{peak}$ & 	0.61111	&	0.8213	&	0.70042	&	0.81528	*\\
			$t_{c1\%}$ and $h_{peak}$ & 	0.54	&	0.81542	&	0.64849	&	0.81042	\\
			$t_{c2\%}$ and $h_{peak}$ & 	0.73333	&	0.82487	&	0.77601	&	0.82222	*\\
			$t_{c3\%}$ and $h_{peak}$ &	0.4986	&	0.82247	&	0.62045	&	0.80903	\\
			$t_{c4\%}$ and $h_{peak}$ & 	0.53611	&	0.82099	&	0.64692	&	0.81111	\\
			$t_{c5\%}$ and $h_{peak}$ & 	0.54444	&	0.82066	&	0.65434	&	0.81181	\\
			$t_{c1\%}$ and $t_{peak}$ & 	0.405	&	0.82146	&	0.54198	&	0.79861	\\
			$t_{c2\%}$ and $t_{peak}$ & 	0.38937	&	0.82222	&	0.52834	&	0.79514	\\
			$t_{c3\%}$ and $t_{peak}$ & 	0.41234	&	0.82288	&	0.54922	&	0.79861	\\
			$t_{c4\%}$ and $t_{peak}$ & 	0.41671	&	0.82172	&	0.55274	&	0.8	\\
			$t_{c5\%}$ and $t_{peak}$ & 	0.46824	&	0.82644	&	0.59694	&	0.80486	\\
			$h_{peak}$ and $t_{peak}$ & 	0.60214	&	0.82118	&	0.69256	&	0.81458	*\\
			$t_{c1\%}$ and $h_{peak}$ and $t_{peak}$ & 	0.36449	&	0.82143	&	0.50442	&	0.79097	\\
			$t_{c1\%}$ and $h_{peak}$ and $t_{peak}$ & 	0.33394	&	0.82022	&	0.47447	&	0.78472	\\
			$t_{c1\%}$ and $h_{peak}$ and $t_{peak}$ & 	0.35772	&	0.82116	&	0.49819	&	0.78958	\\
			$t_{c1\%}$ and $h_{peak}$ and $t_{peak}$ & 	0.38993	&	0.82188	&	0.52849	&	0.79583	\\
			$t_{c1\%}$ and $h_{peak}$ and $t_{peak}$ & 	0.42696	&	0.82541	&	0.56109	&	0.79931	\\

			\bottomrule
		\end{tabular}
	\end{center}
\end{table}

\begin{table}[h!]
	\begin{center}
		\caption{Classification scores based on heat kernel features in the $\neg edge$ partition, averaged across five iterations of stratified 10-fold cross validation with GNB. * denote when all performance measures exceed 60\%.}
		\label{Atab:nonedge_HK_as_is}
		\begin{tabular}{lllll}
			\toprule
			\hline			
			Feature Set & Sensitivity & Specificity & F-Score & Accuracy\\
			\midrule
			$t_{c1\%}$ & 	0.27492	&	0.81126	&	0.40754	&	0.79792	\\
			$t_{c2\%}$ & 	0.0	&	0.80783	&	0.0	&	0.80278	\\
			$t_{c3\%}$ & 	0.0	&	0.80769	&	0.0	&	0.80208	\\
			$t_{c4\%}$ & 	0.0	&	0.80742	&	0.0	&	0.80069	\\
			$t_{c5\%}$ & 	0.66667	&	0.81461	&	0.72975	&	0.81319	*\\
			$h_{peak}$ & 	0.61538	&	0.82909	&	0.70643	&	0.81944	*\\
			$t_{peak}$ & 	0.0	&	0.80783	&	0.0	&	0.80278	\\
			$t_{c1\%}$ and $h_{peak}$ & 	0.45833	&	0.82618	&	0.58852	&	0.80347	\\
			$t_{c2\%}$ and $h_{peak}$ & 	0.53429	&	0.82784	&	0.64916	&	0.8125	\\
			$t_{c3\%}$ and $h_{peak}$ &	0.50745	&	0.82733	&	0.62874	&	0.80972	\\
			$t_{c4\%}$ and $h_{peak}$ & 	0.50667	&	0.82733	&	0.62836	&	0.80972	\\
			$t_{c5\%}$ and $h_{peak}$ & 	0.4549	&	0.82618	&	0.58664	&	0.80347	\\
			$t_{c1\%}$ and $t_{peak}$ & 	0.2858	&	0.8142	&	0.42186	&	0.7875	\\
			$t_{c2\%}$ and $t_{peak}$ & 	0.18	&	0.80902	&	0.26005	&	0.8	\\
			$t_{c3\%}$ and $t_{peak}$ & 	0.0	&	0.80468	&	0.0	&	0.78681	\\
			$t_{c4\%}$ and $t_{peak}$ & 	0.0	&	0.80579	&	0.0	&	0.79236	\\
			$t_{c5\%}$ and $t_{peak}$ & 	0.3254	&	0.81165	&	0.45719	&	0.8	\\
			$h_{peak}$ and $t_{peak}$ & 	0.5419	&	0.82796	&	0.65468	&	0.81319	\\
			$t_{c1\%}$ and $h_{peak}$ and $t_{peak}$ & 	0.40465	&	0.82694	&	0.54285	&	0.79444	\\
			$t_{c1\%}$ and $h_{peak}$ and $t_{peak}$ & 	0.50078	&	0.82769	&	0.62379	&	0.80903	\\
			$t_{c1\%}$ and $h_{peak}$ and $t_{peak}$ & 	0.41731	&	0.82515	&	0.55411	&	0.79792	\\
			$t_{c1\%}$ and $h_{peak}$ and $t_{peak}$ & 	0.46536	&	0.82644	&	0.59537	&	0.80486	\\
			$t_{c1\%}$ and $h_{peak}$ and $t_{peak}$ & 	0.4277	&	0.82589	&	0.56337	&	0.79931	\\

			\bottomrule
		\end{tabular}
	\end{center}
\end{table}

\begin{table}[h!]
	\begin{center}
		\caption{Classification scores based on heat kernel features with GA regressed out in the $\neg edge$ partition, averaged across five iterations of stratified 10-fold cross validation with GNB. * denote when all performance measures exceed 60\%.}
		\label{Atab:nonedge_HK_GA}
		\begin{tabular}{lllll}
			\toprule
			\hline			
			Feature Set & Sensitivity & Specificity & F-Score & Accuracy\\
			\midrule
			$t_{c1\%}$ & 	0.0	&	0.80825	&	0.0	&	0.80278	\\
			$t_{c2\%}$ & 	0.0	&	0.80863	&	0.0	&	0.80694	\\
			$t_{c3\%}$ & 	0.23333	&	0.80956	&	0.31282	&	0.80278	\\
			$t_{c4\%}$ & 	0.53333	&	0.81409	&	0.63582	&	0.81042	\\
			$t_{c5\%}$ & 	0.43333	&	0.81463	&	0.56	&	0.80625	\\
			$h_{peak}$ & 	0.65641	&	0.82959	&	0.73273	&	0.82222	*\\
			$t_{peak}$ & 	0.0	&	0.80876	&	0.0	&	0.80764	\\
			$t_{c1\%}$ and $h_{peak}$ & 	0.52465	&	0.82758	&	0.64067	&	0.81111	\\
			$t_{c2\%}$ and $h_{peak}$ & 	0.58901	&	0.82872	&	0.68838	&	0.81736	\\
			$t_{c3\%}$ and $h_{peak}$ &	0.55619	&	0.82822	&	0.6653	&	0.81458	\\
			$t_{c4\%}$ and $h_{peak}$ & 	0.54095	&	0.82796	&	0.65425	&	0.81319	\\
			$t_{c5\%}$ and $h_{peak}$ & 	0.48824	&	0.82695	&	0.61385	&	0.80764	\\
			$t_{c1\%}$ and $t_{peak}$ & 	0.05	&	0.8069	&	0.07641	&	0.79583	\\
			$t_{c2\%}$ and $t_{peak}$ & 	0.0	&	0.80634	&	0.0	&	0.79514	\\
			$t_{c3\%}$ and $t_{peak}$ & 	0.28939	&	0.81202	&	0.41757	&	0.79722	\\
			$t_{c4\%}$ and $t_{peak}$ & 	0.39717	&	0.81578	&	0.53284	&	0.80278	\\
			$t_{c5\%}$ and $t_{peak}$ & 	0.35495	&	0.81529	&	0.49269	&	0.79792	\\
			$h_{peak}$ and $t_{peak}$ & 	0.56498	&	0.82834	&	0.67132	&	0.81528	\\
			$t_{c1\%}$ and $h_{peak}$ and $t_{peak}$ & 	0.44667	&	0.82592	&	0.57914	&	0.80208	\\
			$t_{c1\%}$ and $h_{peak}$ and $t_{peak}$ & 	0.46601	&	0.82644	&	0.59571	&	0.80486	\\
			$t_{c1\%}$ and $h_{peak}$ and $t_{peak}$ & 	0.4454	&	0.82579	&	0.57694	&	0.80139	\\
			$t_{c1\%}$ and $h_{peak}$ and $t_{peak}$ & 	0.42341	&	0.82527	&	0.55899	&	0.79861	\\
			$t_{c1\%}$ and $h_{peak}$ and $t_{peak}$ & 	0.42152	&	0.82528	&	0.55789	&	0.79861	\\
			\bottomrule
		\end{tabular}
	\end{center}
\end{table}

\begin{table}[h!]
	\begin{center}
		\caption{Classification scores based on heat kernel features with SA regressed out in the $\neg edge$ partition, averaged across five iterations of stratified 10-fold cross validation with GNB. * denote when all performance measures exceed 60\%.}
		\label{Atab:nonedge_HK_SA}
		\begin{tabular}{lllll}
			\toprule
			\hline			
			Feature Set & Sensitivity & Specificity & F-Score & Accuracy\\
			\midrule
			$t_{c1\%}$ & 	0.18333	&	0.80936	&	0.28151	&	0.79722	\\
			$t_{c2\%}$ & 	0.0	&	0.80836	&	0.0	&	0.80556	\\
			$t_{c3\%}$ & 	0.0	&	0.80742	&	0.0	&	0.80069	\\
			$t_{c4\%}$ & 	0.0	&	0.80823	&	0.0	&	0.80486	\\
			$t_{c5\%}$ & 	0.5	&	0.81005	&	0.48219	&	0.80764	\\
			$h_{peak}$ & 	0.61538	&	0.82909	&	0.70643	&	0.81944	*\\
			$t_{peak}$ & 	0.0	&	0.80796	&	0.0	&	0.80347	\\
			$t_{c1\%}$ and $h_{peak}$ & 	0.44133	&	0.82579	&	0.57474	&	0.80139	\\
			$t_{c2\%}$ and $h_{peak}$ & 	0.5419	&	0.82796	&	0.65468	&	0.81319	\\
			$t_{c3\%}$ and $h_{peak}$ &	0.52	&	0.82759	&	0.63854	&	0.81111	\\
			$t_{c4\%}$ and $h_{peak}$ & 	0.52095	&	0.82758	&	0.63898	&	0.81111	\\
			$t_{c5\%}$ and $h_{peak}$ & 	0.47059	&	0.82657	&	0.59973	&	0.80556	\\
			$t_{c1\%}$ and $t_{peak}$ & 	0.15	&	0.8082	&	0.24053	&	0.78681	\\
			$t_{c2\%}$ and $t_{peak}$ & 	0.0	&	0.80729	&	0.0	&	0.8	\\
			$t_{c3\%}$ and $t_{peak}$ & 	0.0	&	0.80482	&	0.0	&	0.7875	\\
			$t_{c4\%}$ and $t_{peak}$ & 	0.0	&	0.80551	&	0.0	&	0.79097	\\
			$t_{c5\%}$ and $t_{peak}$ & 	0.1119	&	0.80736	&	0.18024	&	0.79375	\\
			$h_{peak}$ and $t_{peak}$ & 	0.54952	&	0.82809	&	0.66021	&	0.81389	\\
			$t_{c1\%}$ and $h_{peak}$ and $t_{peak}$ & 	0.38672	&	0.8241	&	0.52577	&	0.79236	\\
			$t_{c1\%}$ and $h_{peak}$ and $t_{peak}$ & 	0.49556	&	0.82708	&	0.61929	&	0.80833	\\
			$t_{c1\%}$ and $h_{peak}$ and $t_{peak}$ & 	0.44199	&	0.82579	&	0.57508	&	0.80139	\\
			$t_{c1\%}$ and $h_{peak}$ and $t_{peak}$ & 	0.44032	&	0.8258	&	0.57415	&	0.80139	\\
			$t_{c1\%}$ and $h_{peak}$ and $t_{peak}$ & 	0.41263	&	0.82502	&	0.55005	&	0.79722	\\

			\bottomrule
		\end{tabular}
	\end{center}
\end{table}

\begin{table}[h!]
	\begin{center}
		\caption{Classification scores based on standard network features, averaged across five iterations of stratified 10-fold cross validation with GNB. CPL = characteristic path length, eBC = Average edge betweenness centrality, CC = Average clustering coefficient, $G_{eff}$ = Global efficiency.}
		\label{Atab:standard_measures}
		\begin{tabular}{lllll}
			\toprule
			\hline			
			Feature Set & Sensitivity & Specificity & F-Score & Accuracy\\
			\midrule		
			Average CPL, eBC, CC, $G_{eff}$ & 0.475 & 0.838 & 0.606 & 0.798\\
			$Edge$ CPL, eBC, CC, $G_{eff}$ & 0.222 & 0.822 & 0.350 & 0.562\\
			$\neg Edge$ CPL, eBC, CC, $G_{eff}$ & 0.450 & 0.837 & 0.585 & 0.792\\
			$Edge$ CPL, $\neg Edge$ CPL, eBC, CC, $G_{eff}$ & 0.416 & 0.837 & 0.556 & 0.781\\
			\bottomrule	
			
		\end{tabular}
	\end{center}
\end{table}

\end{document}